\documentclass[preprint2]{aastex}

\usepackage{graphicx}
\usepackage[authoryear]{}

\newcommand\msun{M$_{\odot}$}

\begin{document}

\title{Pulsations and Hydrodynamics of Luminous Blue Variable Stars\footnote{Updated, corrected, and LaTeX typeset version of \citet{guzik12} published in Astronomical Review, Vol. 7, p. 3, July 2012.}}

\author{J.A. Guzik\altaffilmark{a} and C. C. Lovekin\altaffilmark{b,c}}

\altaffiltext{a}{Theoretical Design Division, Los Alamos National Laboratory, XTD-NTA, MS T-086, Los Alamos, NM  87545 USA}
\altaffiltext{b}{Theoretical Division, Los Alamos National Laboratory, T-2, MS B-227, Los Alamos, NM  87545 USA}
\altaffiltext{c}{Physics Department, Mount Allison University, Sackville, New Brunswick, Canada}

\begin{abstract}
The Luminous Blue Variable stars exhibit behavior ranging from light curve `microvariations' on timescales of tens of days, to `outbursts' accompanied by mass loss of $\sim10^{-3 }$ \msun\ occurring decades apart, to `giant eruptions' such as seen in Eta Carinae ejecting one or more solar masses and recurring on timescales of centuries.  Here we review the work of the Los Alamos group since 1993 to investigate pulsations and instabilities in massive stars using linear pulsation models and non-linear hydrodynamic models.  The models predict pulsational variability that may be associated with the microvariations.  Using a nonlinear pulsation hydrodynamics code with a time-dependent convection treatment, we show that, in some circumstances, the Eddington limit is exceeded periodically in the pulsation driving region of the stellar envelope, accelerating the outer layers, and perhaps initiating mass loss or LBV outbursts.  We discuss how pulsations and mass loss may be responsible for the location of the Humphreys-Davidson Limit in the H-R diagram.  The `giant eruptions', however, must involve much deeper regions in the stellar core to cause such large amounts of mass to be ejected.  We review and suggest some possible explanations, including mixing from gravity modes, secular instabilities, the epsilon mechanism, or the SASI instability as proposed for Type II supernovae.  We outline future work and required stellar modeling capabilities to investigate these possibilities.
\end{abstract}


\section{Introduction}
\label{intro}

The Luminous Blue Variables (LBVs) are a class of high-mass stars that undergo eruptive events accompanied by mass loss.  Many have been discovered in our galaxy, but many are also found in the Large Magellanic Cloud \citep[e.g.][]{wolf00}.  The LBVs show three different types of variability.  In one form, the LBV seems to traverse the H-R diagram at a constant luminosity, but with decreasing effective temperature/increasing radius and increase of 1-2 visual magnitudes during the outburst \citep{puls08}.  This type of variability is known as S Doradus (SD) variability, and is seen as either a short SD phase, with a timescale less than 10 years, or a long SD phase, with a timescale greater than 20 years.  

Another class, or perhaps a separate kind of eruption in the same star, is accompanied by both luminosity and mass increases, with ejection of several solar masses of material, and brightening by 1-3 bolometric magnitudes, and many visual magnitudes.  These stars, named the `supernova impostors' \citep[e.g.,][]{smith11,vandyk12}, may have recurrences of these `giant eruptions' on timescales of centuries.  In our galaxy, the most well-known examples of the giant eruptions are the LBVs P Cyg and Eta Car.  Several candidates have also been detected in the Large Magellanic Cloud \citep{wolf00} and external galaxies \citep[e.g., M33,][]{clark12}.

A third kind of variation in the Luminous Blue Variables is the `microvariations' in luminosity, of a few tenths of a stellar magnitude.  These microvariations appear stochastic in nature, but have irregular periods of order of weeks to months \citep{abolmasov11}.  See also \cite{vink12} for a recent review of LBV properties and research questions.  

The first and second types of variability recently have been proposed as an explanation for various features seen in supernova (SN) light curves.  For example, \citet{kotak06} have proposed that SNe which show quasi-sinusoidal variation in their radio light curve have LBVs undergoing the SD type variations as progenitors.  As the mass-loss rates appear to change during a SD outburst \citep[e.g.,][]{stahl01,stahl03}, the resulting circumstellar medium varies in density.  LBVs as progenitors of SNe have also been proposed as the progenitors of Type IIn SNe or superluminous SN \citep[e.g.][]{smith10}.

The evolutionary status of LBVs is also uncertain.  It has long been accepted that massive O stars become LBVs before losing their outer layers to become Wolf-Rayet (WR) stars, and subsequently SN \citep[e.g.,][]{chiosi86}.  Stars will lose mass through radiatively driven winds throughout these phases, and it has been shown that some LBV envelopes have enriched He abundances and evidence of CNO-cycle processing indicating that outer layers have been removed to expose material that had previously experienced hydrogen burning \citep{najarro97,najarro01}.  However, with recent downward revisions of theoretical mass-loss rates, it seems that radiatively driven mass loss cannot remove enough material to match the observed masses of WR stars (about 20 \msun).  If most or all massive stars end as WR stars, the most massive stars may need a series of giant eruption events during an LBV phase to reach the masses of typical WR stars \citep{smith06}.    

The LBVs are cooler than the core He-burning WR stars, which have been stripped of their outer layers, presumably during an earlier LBV phase.  The most luminous LBVs are hotter and brighter than the Humphreys-Davidson Limit \citep[H-D Limit,][]{HD} in the H-R diagram (see Fig.\,\ref{evolvedmodels}), and may be on their first crossing of the H-R diagram in a shell H-burning or core He-burning phase.   At lower initial masses and luminosities, stars are found below the Humphreys-Davidson limit and across the width of the H-R diagram as blue, yellow, and red supergiants.  At present, the physical origin of the Humphreys-Davidson limit is not understood, and the highest initial mass which can evolve to the red supergiant region is determined observationally.

While much analytical work, atmosphere and wind modeling, and observational work has been assembled to understand and categorize LBV phenomena, and some explanations of the origin of LBV winds and outbursts have been proposed, to our knowledge very little research has been published in the refereed literature on the role of stellar pulsations in producing LBV microvariations or initiating outbursts or winds. In addition, a pulsational instability or some other hydrodynamic instability deep in the core must be responsible for initiating LBV Ôgiant eruptionsÕ.

Here we review the work of the group at Los Alamos National Laboratory from 1993 to the present that has before only been summarized briefly in abstracts or published in conference proceedings.\footnote{Since the original publication of this paper, work in this area by Lovekin \& Guzik has been submitted for publication in MNRAS.}  (These conference presentations are designated by asterisks in the References.)  We focus on linear stellar pulsation modeling using 1-D stellar evolution models including mass loss, and 1-D nonlinear hydrodynamic envelope models using the Dynstar Lagrangian stellar pulsation hydrodynamics code \citep{ostlie82,ostlie90,cox83,cox90,coxostlie93} which includes a time-dependent convection treatment.  We also suggest some possible mechanisms for initiating LBV giant eruptions.  We hope that this review will motivate work by those who have more advanced computational tools to further investigate instabilities that may be responsible for LBV episodic mass loss and giant eruptions.

For a recent comprehensive overview of the properties of Eta Car and other known LBV stars, we recommend the review articles in the volume edited by Davidson and Humphreys, Eta Carinae and the Supernova Impostors, vol. 384 of the Springer Astrophysics and Space Science Library published in 2012.

\section{Pulsations and LBV Outbursts}
\label{PulsationsOutbursts}

\begin{figure}
\includegraphics[width=\columnwidth]{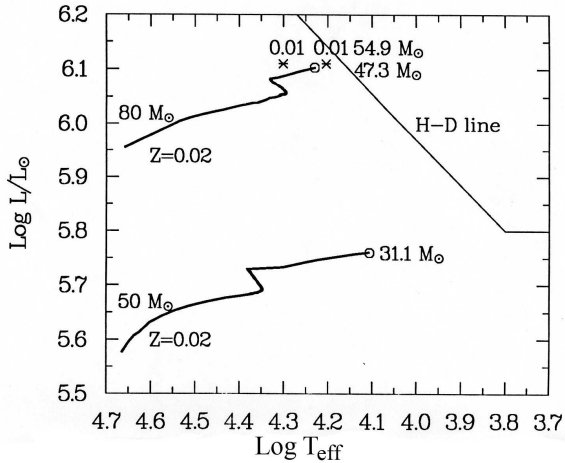}
\caption{Evolution tracks and final masses after mass loss for models of initial mass 80 and 50 \msun\ that were analyzed for stellar pulsations (end-point circles and crosses.) The models lie in the region of the H-R diagram where LBV instabilities originate, to the left of the Humphreys-Davidson limit.  The final mass of the Z=0.01 model (after mass loss) is 54.9 \msun.}
\label{evolvedmodels}
\end{figure}

\begin{figure}
\center
\includegraphics[width=\columnwidth]{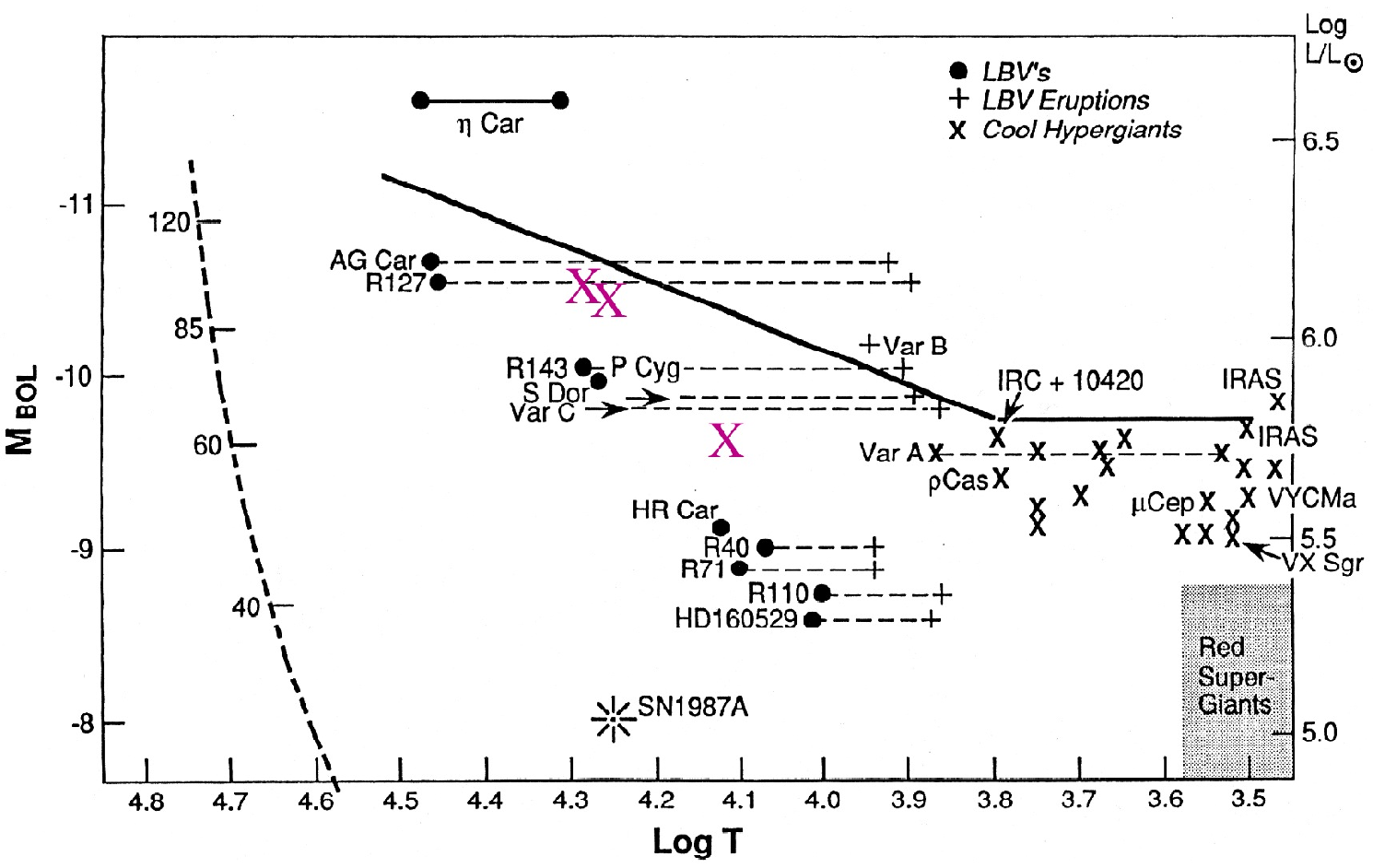}
\caption{Location of models (Red Xs) on H-R diagram from \citep{HD}.}
\label{HDLimit}
\end{figure}

To investigate LBV models, we began in the mid 1990s by evolving some high-mass models using an updated version of the \citet{iben63,iben65a,iben65b} stellar evolution code including mass loss \citep{brunish82a, brunish82b}.   The code included the analytical \citet{stellingwerf75a,stellingwerf75b} fit to the Cox-Tabor opacity tables that was adjusted to roughly calibrate the opacities to Lawrence Livermore OPAL opacities \citep{OPAL96}.  These new higher opacities had recently been shown to explain many problems in stellar modeling, including the Cepheid mass problem, and the cause of $\beta$ Cep star pulsations \citep{rogers94}.   We evolved 50 \msun\  and 80 \msun\ models using the parameterized mass-loss formula of \citet{mlr90} intended to generally fit observed mass-loss rates across the upper H-R diagram.  We adjusted this prescription to increase the mass loss more rapidly with increasing radius, as we wanted to develop models that had a variety of envelope helium abundance enrichments when they arrived in the LBV instability region on their first crossing of the H-R diagram. Figure \ref{evolvedmodels} shows our evolution tracks for 50 and 80 \msun\ models with metallicitiy Z=0.02, appropriate for galactic LBVs.  The two crosses show the endpoints of an 80 \msun\ evolution sequence at Z = 0.01, a metallicity chosen to approximate the lower metallicity of the Large Magellanic Cloud where many LBV stars are found.  Figure \ref{HDLimit} shows the endpoints of the evolution tracks on the H-R diagram of \citet{HD} in relation to many well-known LBV stars and the H-D limit.

We then tested these models for pulsational instabilities using a linear nonadiabatic code developed by \citet{pesnell90}. The models show that pulsations (both radial and nonradial) are driven by the `kappa-gamma' mechanism \citep[see, e.g., J.P.][]{cox80} due to ionization of Fe-group elements in the stellar envelope at $\sim$200,000 K.  The modes are highly nonadiabatic because of the low density and high luminosity of these stars, and the periods can be radically different and even not able to be associated with the corresponding modes found in an adiabatic analysis, motivating them to be called `strange modes' in the literature \citep[see, e.g.,][]{glatzel99a,glatzel99,glatzel94,glatzelmehren96}.  The predicted linear growth rates of strange modes are as large as several hundred percent/period, much higher than the 0.001/period typical of most pulsating variables of lower mass.  Low-mass evolved stars, such as Miras, are also known to have large growth rates \citep[e.g.,][]{coxostlie93}.

We then used the nonlinear hydrodynamics code Dynstar \citep{ostlie82,ostlie90,cox83,coxostlie93} to examine the nonlinear hydrodynamic behavior of these models. The Dynstar code was updated to include the same Stellingwerf opacity fit adjusted to the OPAL opacities used in the evolution modeling. In the hydro code, we construct envelope-only models with 60-120 zones distributed from above the photosphere down to a few million K.  The envelopes of these stars are very tenuous, so our LBV models include 95\% of the star's radius, but only 0.2\% of the mass!  Table \ref{Table1} summarizes the linear pulsation model results for the models that were used as input for the hydrodynamic analysis.  We initiated the pulsations with radial velocity amplitude 1 km/sec outward in the radial mode (Fundamental, 1st Harmonic, or 2nd Harmonic) that we found to be most unstable from the linear analysis (when practical).

The unique feature of the Dynstar code that proved essential to explaining the behavior of the models is a time-dependent treatment of convection.  This treatment is a modification of mixing-length theory \citep{MLT} and allows for the lag in adjustment of the convective flux that is changing dynamically with the pulsation.  Spatial averaging over a few zones allows for convective effects to communicate with nearby zones.  The time-dependent convection treatment is described in more detail in Section \ref{tdc} below.

\begin{figure*}
\center
\includegraphics[width=0.7\columnwidth]{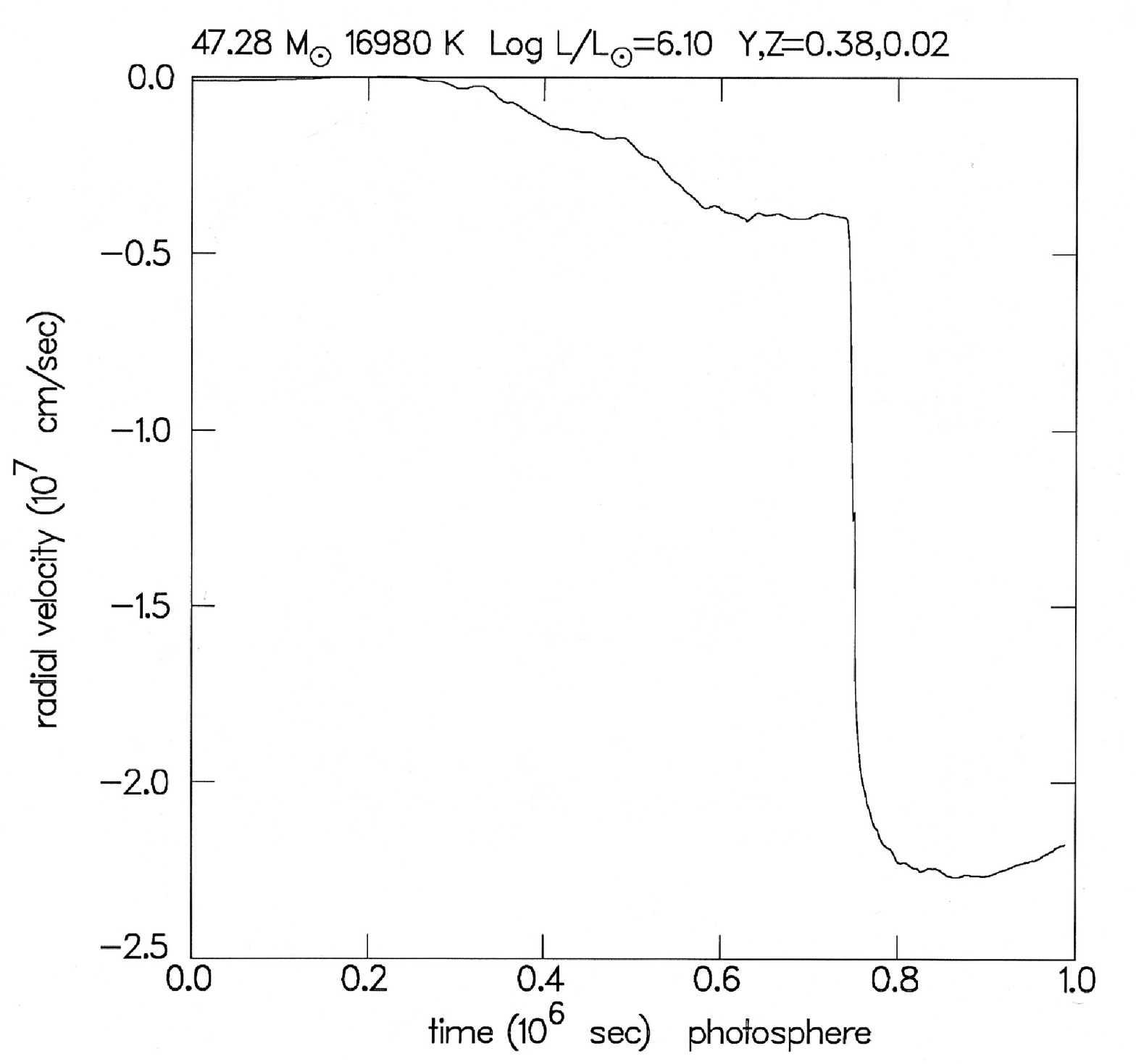}
\includegraphics[width=0.7\columnwidth]{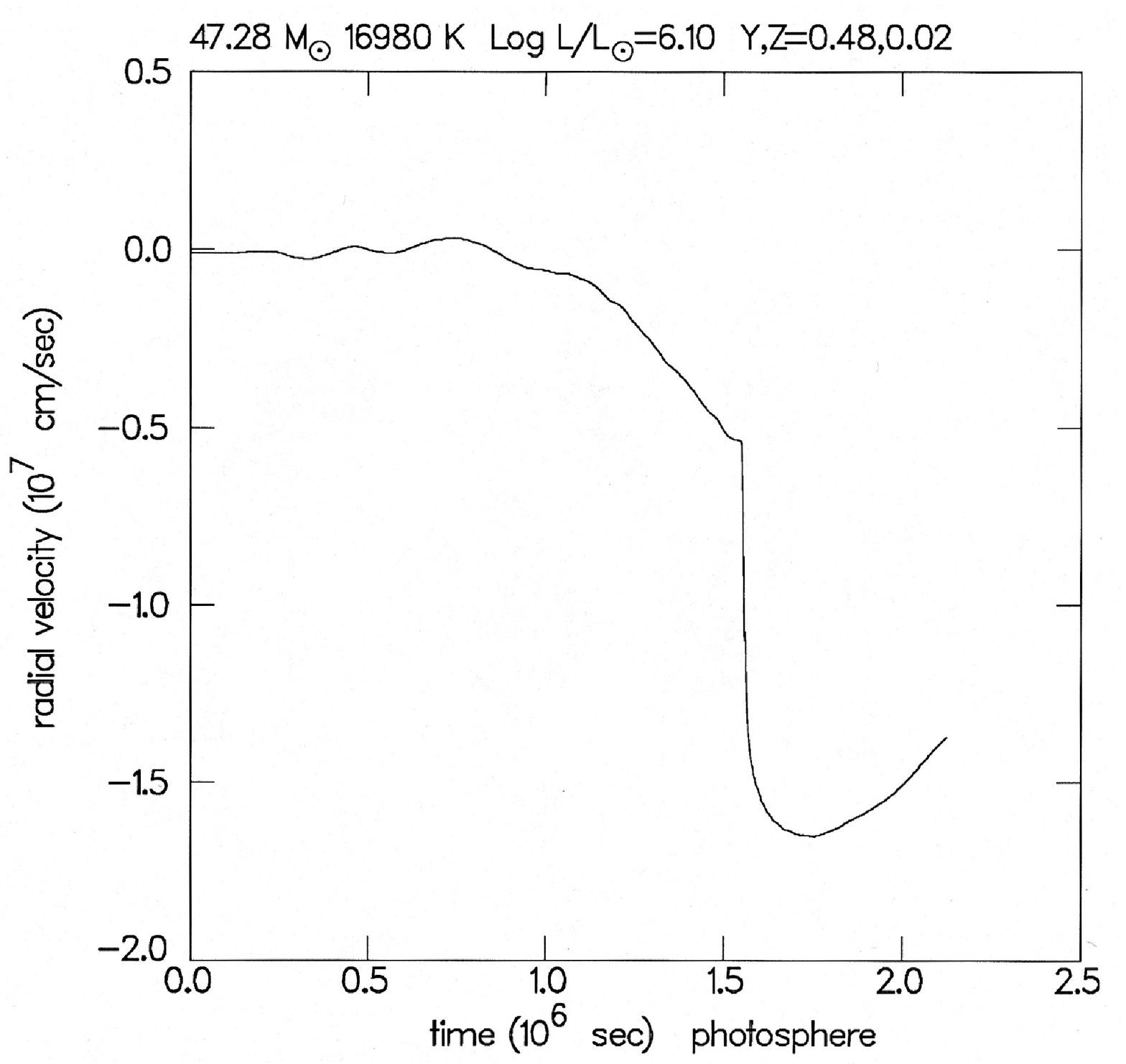}
\includegraphics[width=0.7\columnwidth]{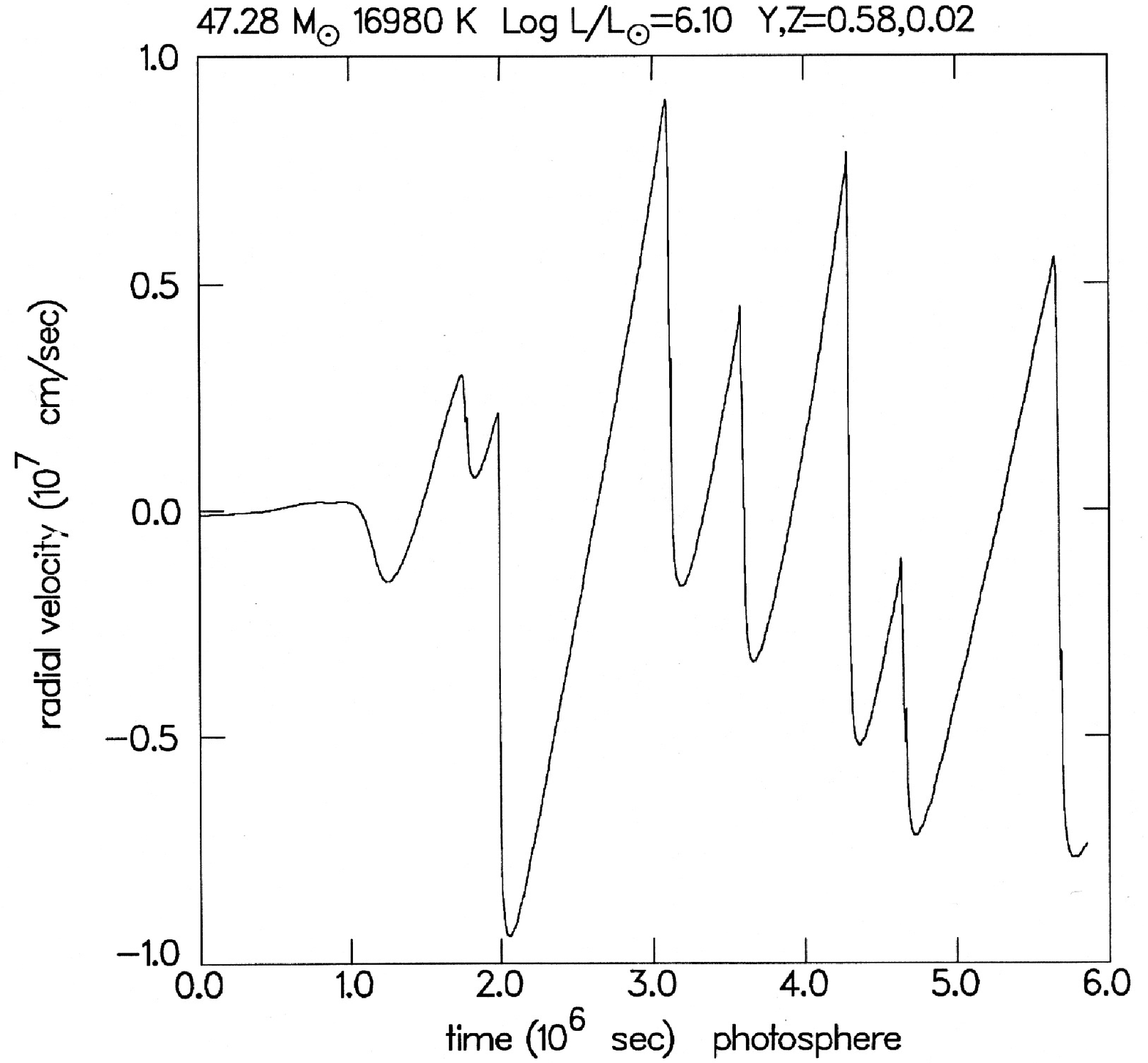}
\caption{ Photospheric radial velocity vs. time for envelope models with helium mass fraction Y=0.38 (left), 0.48 (center),  and 0.58 (right) and Z=0.02.  Negative radial velocity is outward flow.  The Y=0.38 and 0.48 models show large abrupt increase in outward radial velocity.  For comparison, the escape velocity for a non-rotating model is 370 km/sec (3.7 $\times$ 10$^{7}$ cm/sec).  An increase in Y abundance (to 0.58) stabilizes ``outburst'' behavior.}
\label{Fig3abc}
\end{figure*}

\begin{figure*}
\center
\includegraphics[width=0.7\columnwidth]{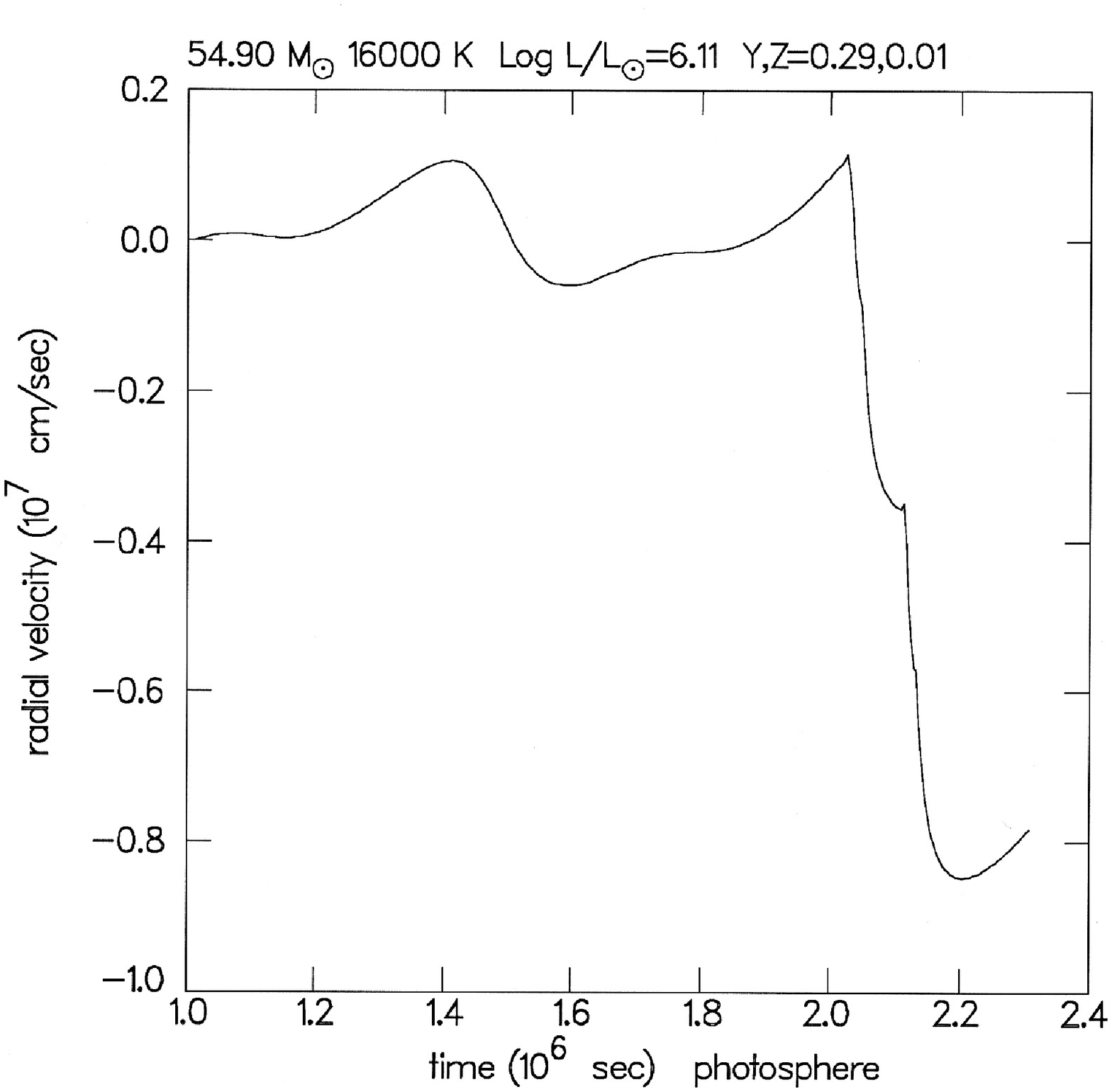}
\includegraphics[width=0.7\columnwidth]{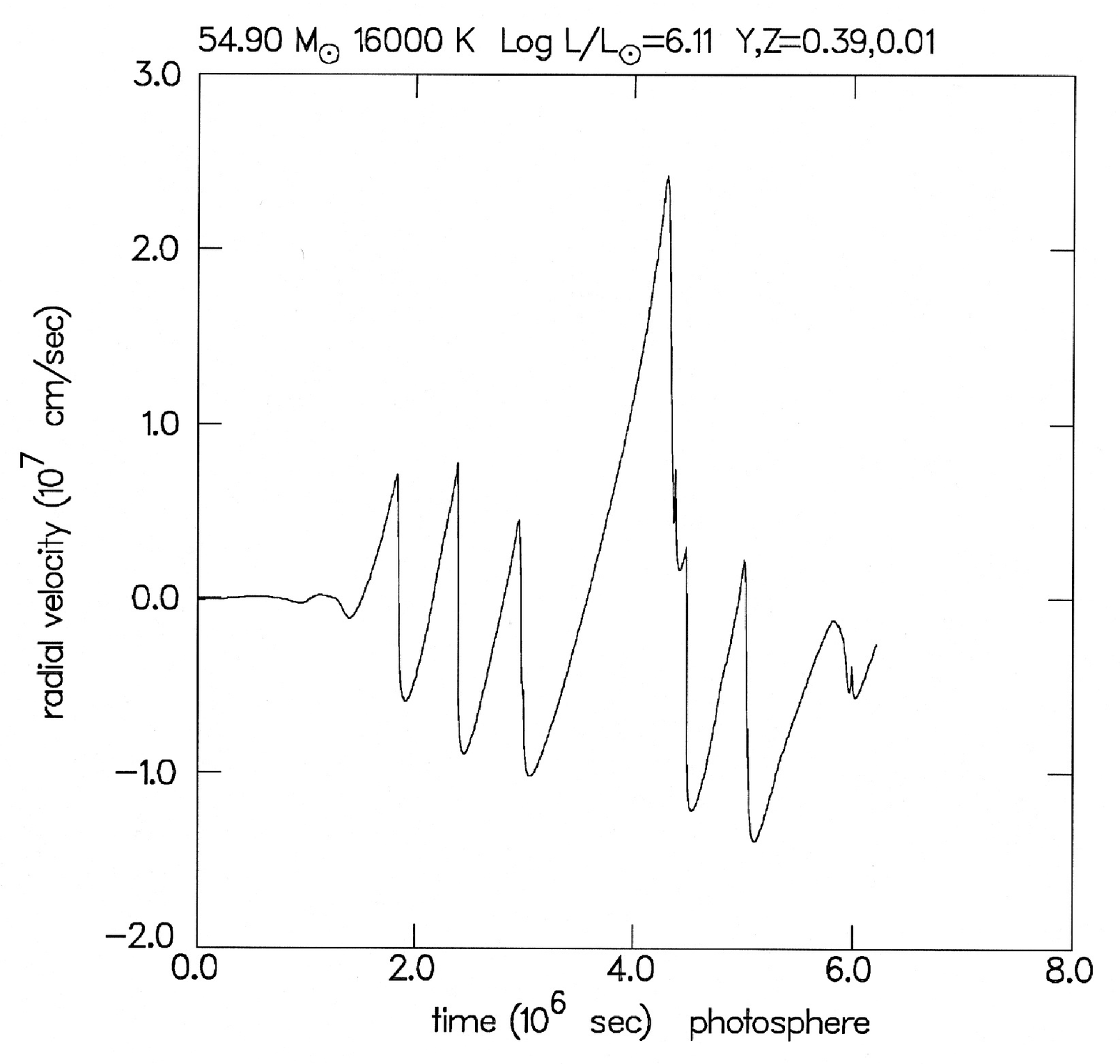}
\includegraphics[width=0.7\columnwidth]{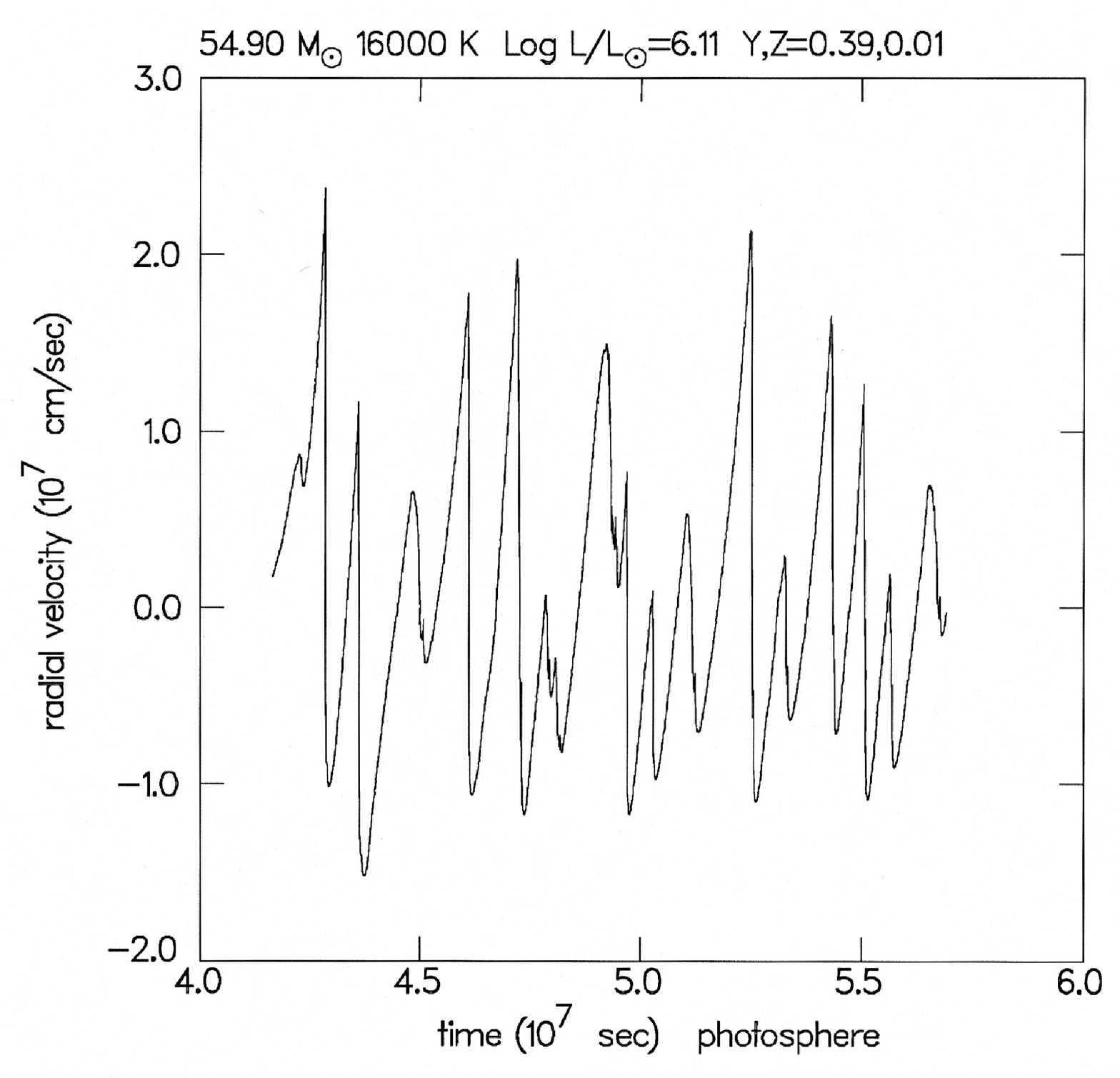}
\caption{Photospheric radial velocity vs. time for envelope models with Y=0.29 and Y=0.39, and Z=0.01. For Y=0.29,  this lower Z=0.01 model (left) shows a large abrupt increase in radial velocity. For comparison, escape velocity for a non-rotating model is 380 km/sec (3.7 $\times$ 10$^{7}$ cm/sec).  For Y=0.39 (center), the model shows steady multimode pulsations.  The model is continued to a later time (right) where the pulsations are still evident.}
\label{Fig4abc} 
\end{figure*}

\begin{figure*}
\center
\includegraphics[width=0.9\columnwidth]{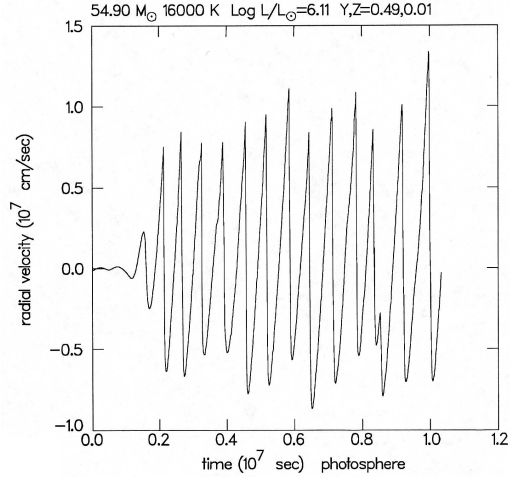}
\caption{Photospheric radial velocity vs.\,time for envelope model with Y=0.49 and Z=0.01. The model shows steady multimode pulsations at lower amplitude than for Y=0.39 models, and period of about 7.6 days.}
\label{Fig5} 
\end{figure*}

\begin{figure*}
\center
\includegraphics[width=0.8\columnwidth]{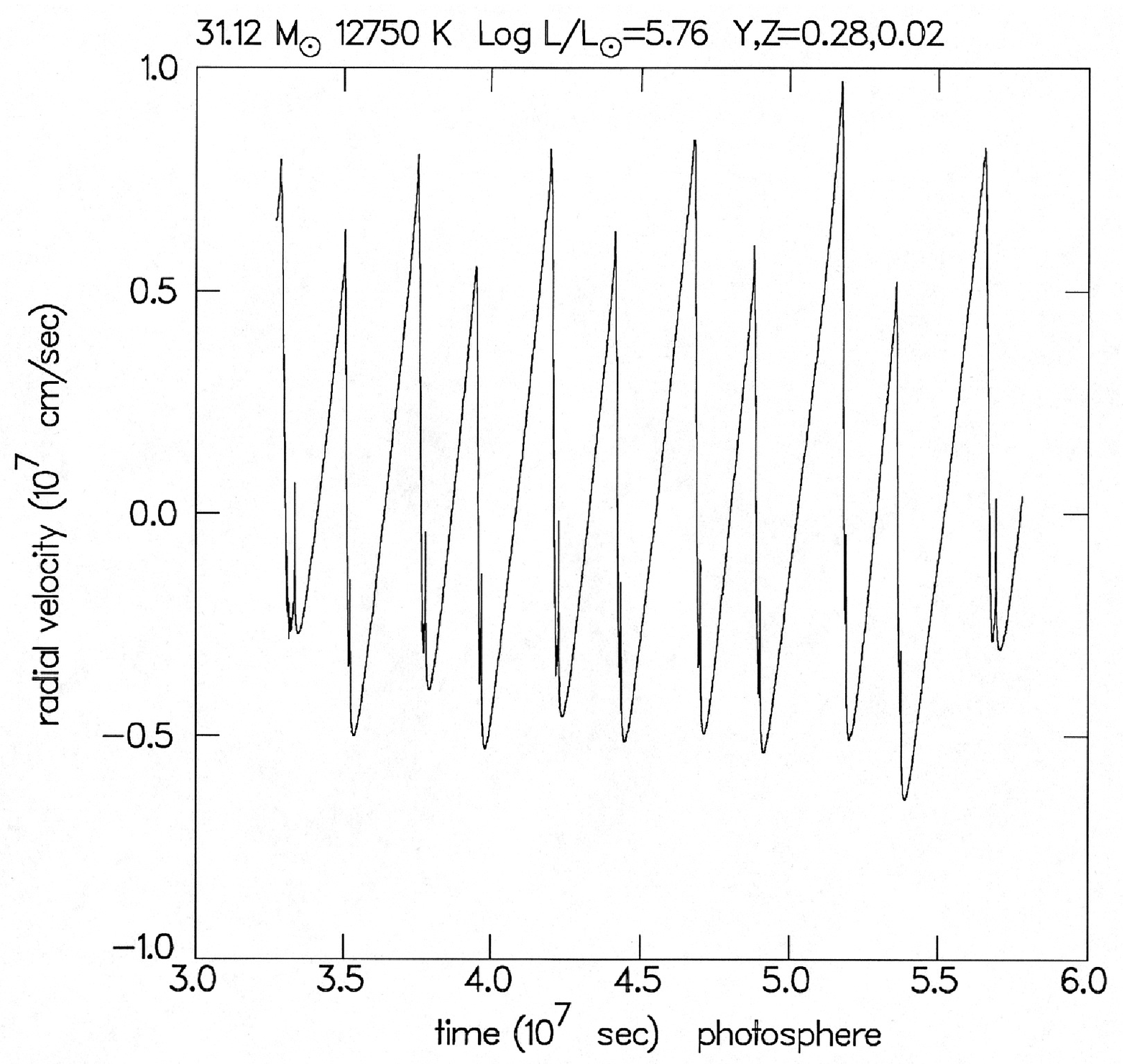}
\includegraphics[width=0.8\columnwidth]{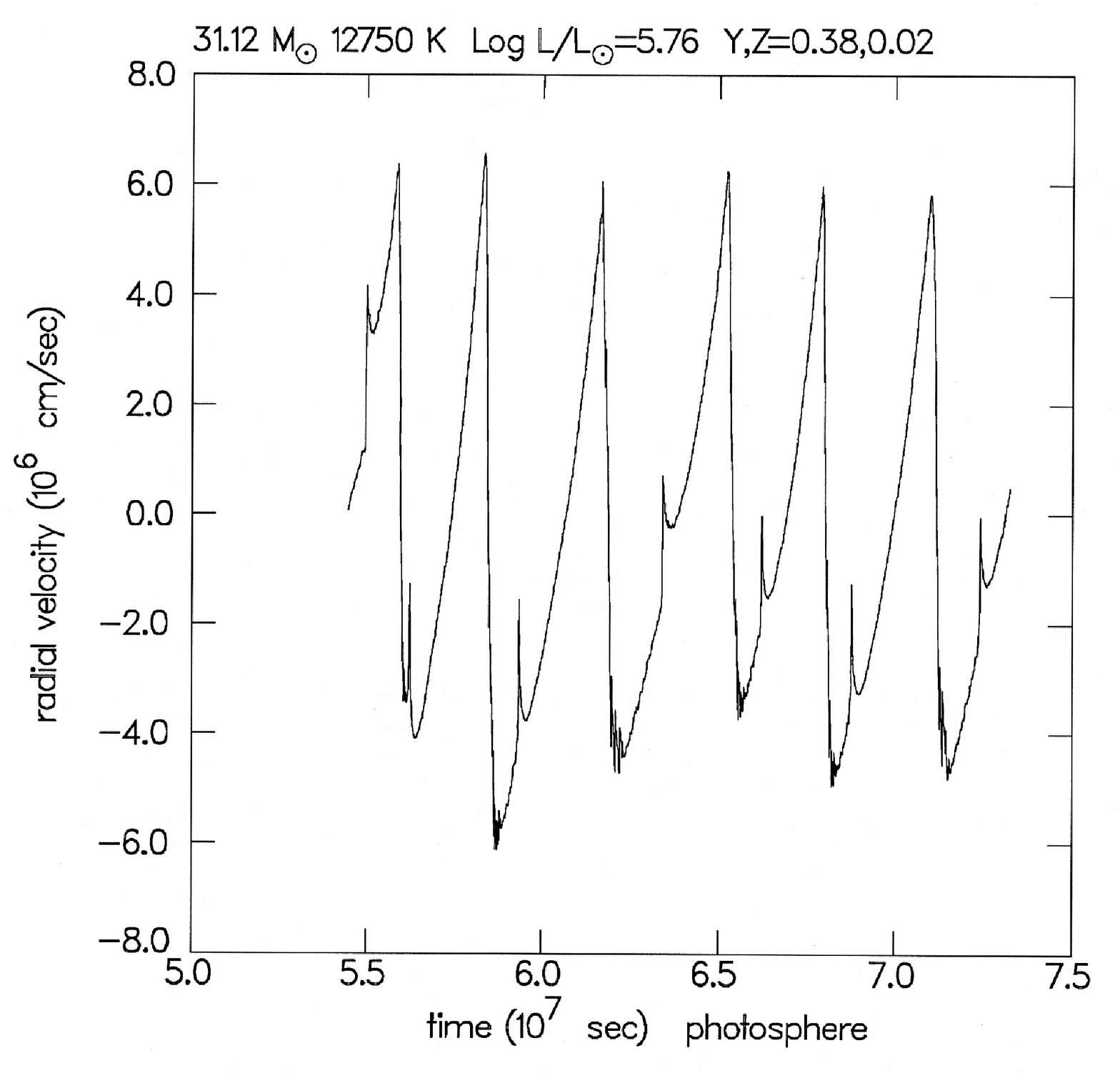}
\caption{Photospheric radial velocity vs.\,time for envelope models with Y=0.28 and 0.38 and Z=0.02. These lower mass and luminosity models exhibit steady pulsations with periods $\sim$26 d for the Y=0.28 model, and $\sim$35 d for the Y=0.38 model.  For comparison, the escape velocity for a non-rotating model is 280 km/sec. (2.8 $\times$ 10$^{7}$ cm/sec)}
\label{Fig6ab}
\end{figure*}

We found in our initial nonlinear study that the pulsation amplitudes are regulated by metallicity and by the envelope helium (Y) abundance \citep{guzik96, guzik99b, guzik04}. Figures \ref{Fig3abc}, \ref{Fig4abc}, \ref{Fig5}, and \ref{Fig6ab} show photospheric radial velocities vs. time, with negative velocities indicating outward flows.   If the He abundance is sufficiently low (the exact value depends on mass and metallicity; see Table \ref{Table2}), the models quickly develop a large outward photospheric velocity.  Layers just below the photosphere continue to pulsate, producing shocks or waves.  These can interact with the photospheric layers to produce an abrupt outward increase in velocity.  The photosphere attempts to recover, but because the hydro code cannot follow mass outflows, and the opacity and equation-of-state (EOS) tables and radiative transfer models are no longer appropriate for the outermost layers well beyond the photosphere, the model cannot be continued further in time.  The final radial velocities we find (about 250 km/sec) are still lower than the escape velocity for a non-rotating model (v$_{esc}$ $\sim$ 370 km/sec).  We associate these abrupt increases in radial velocity with the initiation of an ``outburst''.

\begin{table}
\caption{Linear pulsation model results for models analysed hydrodynamically.}
\begin{tabular}{l}
\\
47.28 \msun\, T$_{\rm eff}$ = 16980 K, Z=0.02\\
High-mass Galactic LBV near H-D Limit\\
\hline
\end{tabular}
\begin{tabular}{lll}
Envelope Y & Linear Period &  Growth rate/ \\
abundance & (days)	& period \\
\hline
0.38 & 9.5 & 0.70 \\
 &	8.6 & 	6.0$^{a}$ (2H)\\
 &	4.4 &1.3\\
0.48 &10.6 & 0.40$^{a}$ (F)\\
 &	33.3 &1.1\\
0.58 & 15.1 & 2.3$^{a}$ (F)\\
	 & 12.2 & 0.80\\
	 & 7.1 & 2.2\\
\\
\end{tabular}

\begin{tabular}{l}
54.9 \msun\ T$_{\rm eff}$ = 16000 K, Z=0.01\\
High-mass LMC LBV near H-D Limit\\
\hline
\end{tabular}
\\
\begin{tabular}{lll}
Envelope Y & Linear Period &  Growth rate/ \\
abundance & (days)	& period \\
\hline
0.29 & 18.0 & 0.40  (F)\\
0.39 & 17.6 & 0.43$^{a}$ (1H)\\
	 & 33.3 & 3.6\\
0.49 &	 28.8 &1.4 (F) \\
\\
\end{tabular}

\begin{tabular}{l}
31.1 \msun\, T$_{\rm eff}$ = 12750 K, Z=0.02\\
Lower-mass LBV below horizontal part\\
of H-D Limit\\
\hline
\end{tabular}
\begin{tabular}{lll}
Envelope Y & Linear Period &  Growth rate/ \\
abundance & (days)	& period \\
\hline
0.28 & 22.4 &	0.96 (F)\\
0.38 & 23.0 & 0.80 (F)\\
\end{tabular}
\tablenotetext{a}{Mode used for hydrodynamic analysis}
\label{Table1}
\end{table}

The increase in envelope He abundance has a stabilizing effect on the models; as the envelope He abundance increases, the outbursts disappear, to be replaced by regular pulsations.  If the He abundance of the model is decreased further, the amplitude of the pulsation decreases. These results are summarized in Table \ref{Table2}, which shows the exact He and metallicity (Z) dependence for each model.

\subsection{The Eddington Limit}
\label{EddLimit}

An important characteristic that determines whether models ``outburst'' (i.e., exhibit an abrupt increase in photospheric radial velocity from which it is difficult to recover) appears to be whether the local radiative luminosity of layers with temperatures greater than 100,000 K exceeds the Eddington Luminosity limit at some point during the pulsation cycle.  The Eddington Luminosity is defined as

\begin{equation}
L_{Edd} = \frac{4\pi GMc}{\kappa}
\end{equation}

\noindent
where $G$ is Newton's gravitational constant, $M$ is the stellar mass, $c$ is the speed of light, and $\kappa$ is the radiative opacity.  When this luminosity is reached, the outward force due to radiation pressure exceeds the force due to gravity.  In a non-pulsating model in hydrostatic equilibrium, if the opacity becomes too high, blocking the emerging radiation, convection turns on to transport the luminosity, so the model stays in hydrostatic equilibrium and avoids exceeding the Eddington limit.   If, in a pulsating model, the convection is allowed to adapt instantaneously, convection will also turn on and prevent the local radiative luminosity from ever exceeding the Eddington luminosity.  However, in the more realistic pulsating model with time-dependent convection, the convection takes some time to turn on during a pulsation cycle, so the Eddington limit can be periodically exceeded.  {\it Including the time dependence of convection is therefore the key to causing the pulsation driving layers to exceed the Eddington limit and initiate outbursts.}

\begin{figure}
\center
\includegraphics[width=\columnwidth]{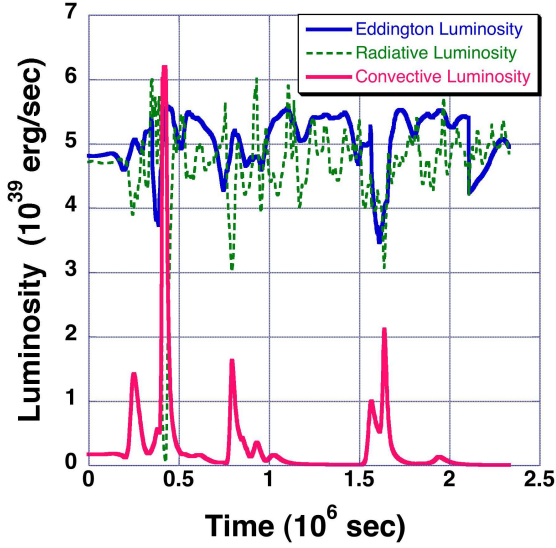}
\caption{Eddington, radiative, and convective luminosity of 47 \msun\ model with T$_{\rm eff}$ = 16,980 K in the driving region of the 60 zone envelope model (T $>$100,000 K).  The radiative luminosity (green) sometimes exceeds the Eddington Limit (blue), until the amount of luminosity carried by convection (red) increases later to transport the excess.}
\label{Edd47}
\end{figure}

\begin{figure}
\center
\includegraphics[width=\columnwidth]{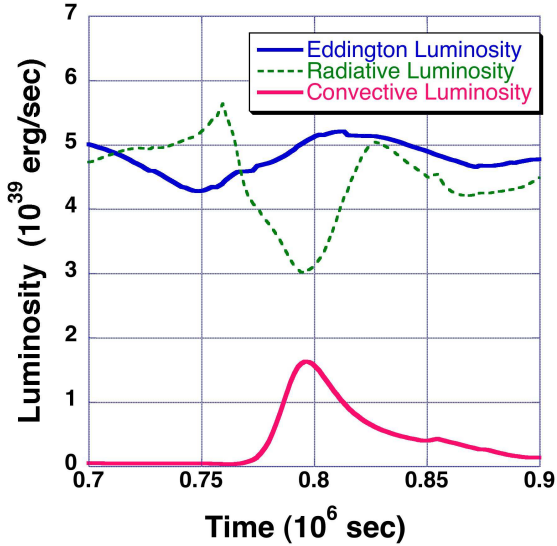}
\caption{Same as Figure \ref{Edd47}, but zooming in on a small time region, where it is clear that the radiative luminosity (green) exceeds the Eddington limit (blue) before convection can turn on to transport the luminosity (red).}
\label{Edd48}
\end{figure}

The nonlinear pulsation results are summarized in Table \ref{Table2}.  For each model, we list whether the model showed an outburst.  If no outburst was present, the maximum velocity is given instead (column 3).  The period of the resulting pulsation (if applicable) is given in column 4.  In column 2, we state whether there were optically thick zones which exceeded the Eddington limit for part of the pulsation cycle.  Outbursts are only triggered if the luminosity exceeds the Eddington luminosity in the driving region, where the temperature is 100,000-200,000 K.  As a result of the delay in the onset of convection, there is a buildup of radiation pressure, which gives a push to the layers above.  Figures \ref{Edd47} and \ref{Edd48} show the Eddington luminosity, convective luminosity and radiative luminosity in the driving region of models of 47 and 31 \msun\ respectively.  The local radiative luminosity periodically exceeds the Eddington luminosity, before dropping sharply when convection turns on.  If there is enough mass in the layers above the region exceeding the Eddington limit, this may result in an outburst.  In many of the models, in the surface zones (near and above the photosphere), the Eddington luminosity is exceeded for at least part of the time, but there is not enough mass in and above these zones to drive an outburst.

\begin{table}
\caption{Summary of nonlinear pulsation results.}
\begin{tabular}{l}
\\
47.28 \msun\, 16980 K, Z=0.02\\
\hline
\end{tabular}
\begin{tabular}{llll}
	Y  &  Optically Thick   &   &  \\
  &   		Zones with	  &  		Outcome  &  Period 	 \\
  &  		L$_{\rm rad}$ $>$ L$_{Edd}$  &     & (days) \\
\hline

	0.38	  &  Yes	  &  	``Outburst''	  &  	n/a\\
	0.48  &  	Yes	  &  	``Outburst''		&    n/a\\
	0.58  &  	Yes		  &  V$_{\rm max}$ 100 km/s	  &  	$\sim$4.6\\
\\
\end{tabular}
\begin{tabular}{l}
54.9 \msun\, 16000 K, Z=0.01 \\
\hline
\end{tabular}
\begin{tabular}{llll}
	Y  &  Optically Thick   &   &  \\
  &   		Zones with	  &  	Outcome	  &  Period 	 \\
  &  		L$_{\rm rad}$ $>$ L$_{Edd}$  &     & (days) \\
    \hline
	0.29	  &  Yes		  &  ``Outburst''	  &  	n/a\\
	0.39  &  	None	  &  	V$_{\rm max}$ 200 km/sec  &  	$\sim$10.5\\
	0.49  &  	None	  &  	V$_{\rm max}$ 100 km/sec	  &  7.6\\
\\
\end{tabular}
\begin{tabular}{l}
\\
31.1 \msun\, 12750 K, Z=0.02\\
\hline
\end{tabular}
\begin{tabular}{llll}
	Y  &  Optically Thick   &   &  \\
  &   		Zones with	  &  	Outcome	  &  Period 	 \\
  &  		L$_{\rm rad}$ $>$ L$_{Edd}$  &     & (days) \\
  \hline

 	0.28  &  	None	  &  	V$_{\rm max}$ 90 km/sec  &  	26\\
	0.38	  &  None	  &  	V$_{\rm max}$ 60 km/sec  &  	35\\
\end{tabular}
\label{Table2}
\end{table}

\subsection{Time dependent convection treatment}
\label{tdc}

The treatment of time-dependent convection used in Dynstar is that outlined by \citet{ostlie90} \citep[see also][]{cox90}.  The treatment is a modification of the \citet{MLT} mixing-length theory.  In this model, the mixing length is the distance a convective eddy can travel before reaching equilibrium with the surrounding medium.  In a static model envelope, the luminosity carried by this convection is given by

\begin{equation}
L_{conv} = 4\pi r^2\frac{4T\rho C_p}{g\ell Q}{v_c^o}^3
\end{equation}
where $r$, $T$, $p$, $\rho$, $C_p$ and $g$ are the local radius, temperature, pressure, density, specific heat and gravity respectively.  The convective velocity, $v_c^o$ is given by
\begin{equation}
v_c^o = \frac{1}{r\sqrt{2}}\frac{g^{1/2}Q^{1/2}\ell}{H_p^{1/2}}f
\end{equation}
where the mixing length, $\ell$, is a variable parameter, typically of order 1.  The pressure scale height, $H_p$ is defined to be $P/\rho g$.  The parameter $Q$ is the logarithmic density gradient
\begin{equation}
Q = \left(-\frac{\partial log \rho}{\partial T}\right )_P
\end{equation}
which is exactly 1 for an ideal gas with constant mean molecular weight.   

The convective velocity, $v_c^o$ is for the static case,  where the parameter $f$ is defined as 
\begin{equation}
f = \left[\sqrt{1  + 4A^2(\nabla-\nabla _{ad})}-1\right]/A
\end{equation}
where 
\begin{equation}
\nabla = \frac{dlogT }{dlogP}
\end{equation}
and 
\begin{equation}
A \equiv \frac{Q^{1/2}C_p\kappa g\rho ^{5/2}\ell ^2}{12\sqrt{2}acP^{1/2}T^3}.
\end{equation}

\noindent
$\nabla-\nabla_{ad}$ is the superadiabatic gradient, $\kappa$ is the local opacity of the material, and $a$ is the radiation constant.

To modify this theory to include time-dependent convection, the convective eddy velocity (and hence the convective luminosity) should respond to changes in the local conditions, but not instantaneously.  The convective eddy velocity is also dependent on effects from neighboring zones.

\begin{figure}
\center
\includegraphics[width=\columnwidth]{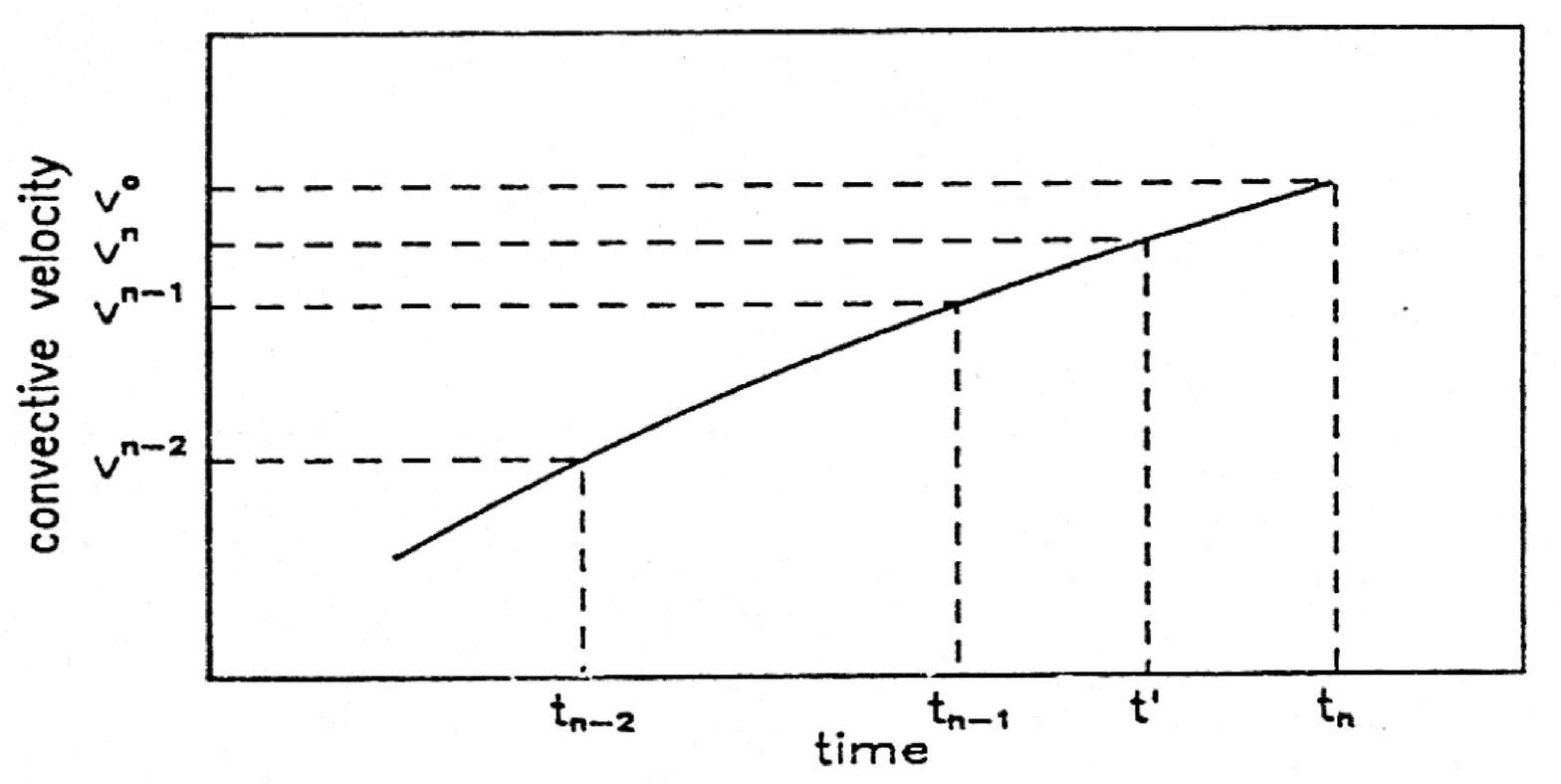}
\caption{The time dependence of the convective velocity.  A three-point Lagrange interpolation polynomial is used to find the time-modulated convective velocity.  Figure from \citet{ostlie90}.}
\label{cvsavg}
\end{figure}

First, we account for the time dependence of the convective velocity, using a quadratic Lagrange interpolation, as illustrated in Fig. \ref{cvsavg}.  This interpolation uses the previous two time steps and the instantaneous value (from the original mixing length theory) to determine the current convective velocity:
\begin{eqnarray}
\nonumber
 v_{c,i}^n = {\frac{(t^{\prime} - t_{n-1})(t^{\prime} - t_{n-2})}{(t_n - t_{n-1})(t_n - t_{n-2})}}v_{c,i}^o \\
 \nonumber
+\, {\frac{(t^{\prime}-t_n)(t^{\prime} - t_{n-2})}{(t_{n-1} - t_n)(t_{n-1} - t_{n-2})}}v_{c,i}^{n-1} \\
  \nonumber
 +\,{\frac{(t^\prime -t_n)(t^{\prime} - t_{n-1})}{(t_{n-2} - t_n)(t_{n-2} - t_{n-1})}}v_{c,i}^{n-2} \\
\end{eqnarray}

where

\begin{equation}
t^{\prime} \equiv t_{n-1} + \tau(t_n - t_{n-1})
\end{equation}
 
and

\begin{equation}
\tau \equiv {\frac{v_{c,i}^n(t_n - t_{n-1})}{\ell}} l_{fac}
\end{equation}

\noindent
for  $ 0 \leq \tau \leq 1 $ and $ l_{fac} \sim$ 1.
 
The quantity $l_{fac}$ is the fraction of a mixing length the convective eddy can travel in one time step.  The $t^{\prime}$ incorporates a phase lag, so there can be some delay before the convective flow begins.  

The convective velocity can also depend on non-local effects, and so this treatment includes a spatial average over neighboring zones.  Zones within one mixing length are assumed to have some contribution to the local conditions.  Incorporating all of these, the convective velocity for zone $i$ at time step $n$ is
\begin{eqnarray}
\nonumber
{\overline{v}_{c,i}^n} = a_{i-1}v_{c,i-1}^{n-1} + a_{i+1}v_{c,i+1}^{n-1} \\
+\,(1-a_{i-1} - a_{i+1})v_{c,i}^n
\end{eqnarray}
where $v_{c,i}^n$ is the convective velocity from the time averaging for the current time step ($n$) and zone ($i$).  The weighting factor $a$ for zone $k$ is given by
\begin{equation}
a_k = a_{fac}\left(1-\frac{|r_k-r_i|}{\ell}\right)
\end{equation}
where $a_k$ constrained to be between 0 and an averaging factor, $a_{fac}$, and $a_{fac} \leq$ 1/3 is a free parameter.
This final form of the convective velocity is then used in the first equation to calculate the time-dependent convective luminosity. 

\subsection{An explanation for the H-D limit?}

For high-mass models near the H-D limit, pulsations in our models grow to large amplitudes ($>$100 km/sec).  The pulsation periods and amplitudes we find are similar to those of observed LBV microvariations, which are typically between 5 and 50 days, with amplitudes of $\sim$0.1 magnitude or less \citep[see, e.g.,][]{vangenderen85,vangenderen90}.

When the deeper (adiabatic) regions of the envelope exceed $L_{Edd}$, an ``outburst'' occurs.
The radial velocity at the photosphere remains negative, and the radii of outer zones monotonically increase during several pulsation cycles.  An increase in envelope helium abundance lowers the opacity in driving regions and appears capable of limiting the pulsation amplitudes and putting an end to the ``outbursts''.

These nonlinear hydro model investigations motivate the following picture for the reason for the H-D limit in the H-R diagram:  Stars of initial mass 
$>$50  \msun\  repeatedly outburst and lose mass until their envelopes are enriched in helium and outbursts are stabilized.  These stars never evolve past the H-D line. Pulsation-driven winds with a high mass-loss rate keep these stars to the blue of H-D line (e.g., S Dor, AG Car).

Stars of initial mass  $<$50 \msun\  pulsate on first crossing of H-R diagram, but do not outburst.  These stars are able to evolve to become yellow and red supergiants.  After they have lost considerably more mass as RSGs and evolve back to the blue, their increased $L/M $ratio may cause them to exceed the Eddington limit and experience outbursts (e.g., HR Car, R40, R71).

In fact, LBVs have been shown to be enriched with He \citep{najarro97}, with Y as high as 0.6.  The effects of this He enrichment on mass-loss rates have been studied by \citet{vink02}, with more helium resulting in higher mass-loss rates at high temperatures, and lower mass-loss rates at low temperatures.  The pulsational stability of massive stars has been studied by \citet{shiode12}, but only in terms of metallicity, not helium composition.  Clearly, there is still both theoretical and observational work to be done in this area to confirm the effects of composition on pulsational stability and outbursts of LBVs.  
  
\subsection{Winds or eruptions?}

Since the envelopes of LBVs are very tenuous, we have shown only in our hydro models that this mechanism can trigger an instability and accelerate shells of 10$^{-4}$ solar masses (the mass of the envelope).  We have not modeled the outflow or winds or recovery from a mass-loss episode.  Perhaps pulsation initiates the flow, and shocks and line-driven winds accelerate it.  Our models may explain pulsation-driven mass loss, or irregular mass-loss rates, modulated by the buildup of pulsation amplitude.  More work is required to follow a mass loss episode and recovery, to determine whether this mechanism could be responsible for the LBV outbursts. 

\section{Hydrodynamics of blue stars of lower initial mass}

Our initial evolution and hydrodynamic models used the Stellingwerf analytical fit to Cox-Tabor opacity tables, augmented by roughly a factor of two to account for increased opacities in newer tables.  In $\sim$1997 we implemented interpolations on the OPAL \citep{OPAL96} opacities, supplemented for layers above the photosphere by the \citet{alexander94,alexander95} tables.   The increased opacities, combined with interpolation in tables caused the  hydrodynamic behavior to become more ragged and less stable.  We were also unable to evolve very massive stars with high luminosity/mass ratio ($M>$80 \msun) including mass loss as the static models were already exceeding the Eddington limit in their outer layers and were not in hydrostatic equilibrium.   To study the hydrodynamic behavior in a more stable regime and facilitate parameter studies, we decided to investigate models of about 20 \msun\ near the top of the $\beta$  Cephei instability strip on the main sequence \citep[see also][]{guzikaustin05}. Note that some comparable nonlinear hydrodynamic calculations were carried out by \citet{dorfi00} for a 30 \msun\ model.

The opacity of elements near Fe (atomic number Z$\sim$26) was found to be important for solving the Cepheid mass problem \citep{rogers94}, and also to explain the pulsations of two classes of nonradially pulsating main-sequence stars, the slowly pulsating B (SPB) stars and the $\beta$ Cephei stars \citep{pamyatnykh02}.   The $\beta$ Cephei and SPB stars are burning hydrogen in their core, and are in the main-sequence phase of stellar evolution.
SPB and $\beta$  Cephei stars pulsate in both radial and nonradial low-amplitude pulsations.
The modulation of radiation due to ionization of Fe-group elements (Fe, Co, Ni) at temperatures in the envelope around 200,000 K can explain the pulsation driving.
As these stars evolve, they exhaust central hydrogen in their convective core, the core collapses, and the star begins burning hydrogen in a shell (after the `blue hook' on the H-R diagram).  The envelope subsequently expands and the stars become more susceptible to higher amplitude radial pulsation since the envelope is less tightly bound.

The kappa ($\kappa$) or more correctly, ``kappa-gamma'' effect is due to the modulated blocking and releasing of radiation during a pulsation cycle by increasing opacity in ionizing regions \citep[see, e.g., J.P.][]{cox80}. This modulation allows work to be put into increasing the amplitude of a perturbation, until enough energy begins to be dissipated again at some limiting amplitude.  If extra energy is blocked in a layer when the star is smallest (layer is compressed), then it will increase the pressure and give an extra boost to expansion.  If the energy is then released when the star is largest (layer is expanded), this lowers the pressure and removes some support against gravity, so that the layer overshoots its equilibrium radius at compression.

For a mode to be pulsationally unstable, the integrated driving throughout all layers must exceed the radiative damping. Regions that produce pulsating driving by the $\kappa$ effect are where the opacity increases with increasing temperature, as it does in ionizing regions.  Stars are known to pulsate due to the ionization of hydrogen (e.g. Mira red giant variables, DA white dwarfs), helium (e.g. Cepheids, $\delta$ Scuti stars, DB white dwarfs), and iron (e.g. $\beta$  Cephei and Slowly Pulsating B stars).  See \cite{aerts10} for a recent volume on pulsating variable stars.

\begin{figure}
\center
\includegraphics[width=\columnwidth]{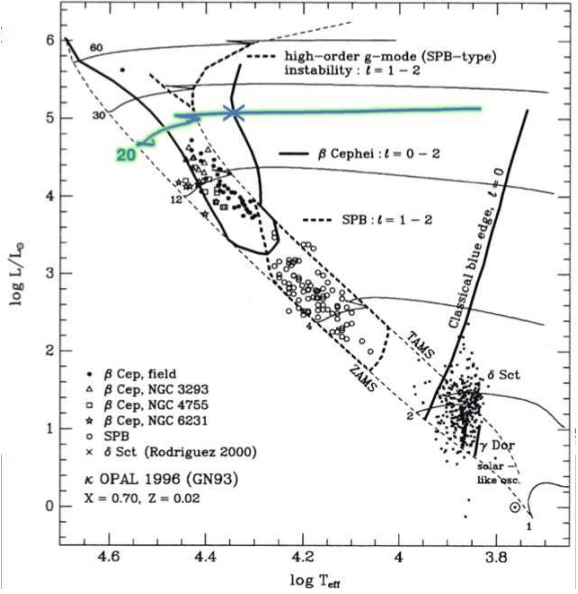}
\caption{H-R diagram showing the main-sequence variables (from Pamyatnykh 2002).    The green evolution track is our 20 \msun\ model superimposed. The green cross marks the position where we tested our model for pulsations.}
\label{Pamyatnykh}
\end{figure}

Figure \ref{Pamyatnykh} shows our evolution track calculated with the Iben code for composition X, Y, Z = 0.70, 0.28, 0.02 using the \citet{GN93} solar element mixture superimposed on the H-R diagram of Pamyatnykh (2002).   The green X marks the location of the T$_{\rm eff}$$\sim$22,000 K models for which we studied the pulsational stability.  For a 20 \msun\ model, the radius increases from 6 R$_{\odot}$ on the zero-age main sequence to 15 R$_{\odot}$ before the `blue hook' at the end of the main sequence, and then to 23 R$_{\odot}$ after the `blue hook'.  Our linear pulsation analysis used a model-building code described by \citet{cox83} that was updated to include the OPAL opacities, and the \citet{pesnell90} nonadiabatic pulsation code with 500 zones, and we used the Dynstar code for the 60-zone envelope hydrodynamic models. 

To investigate the effects of composition and temperature more systematically, we first examined the driving region using linear models, where we could vary one parameter at a time (Y, Z, T$_{\rm eff}$). Figures 11-14 show the results of these tests. 

\begin{figure}
\center
\includegraphics[width=\columnwidth]{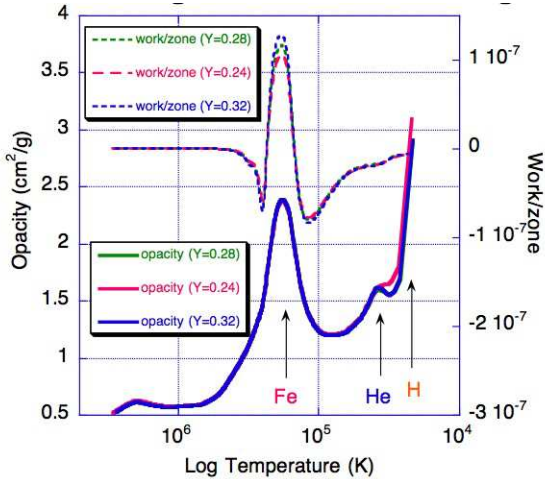}
\caption{Opacity and work per zone for nominal 20 \msun\ model with T$_{\rm eff}$ = 22,000 K in which we varied the envelope Y abundance.  For fixed effective temperature, radius ($\sim$23 R$_{\odot}$), and luminosity, the pulsation driving increases with increasing Y.}
\label{Fig11}
\end{figure}

\begin{figure}
\center
\includegraphics[width=\columnwidth]{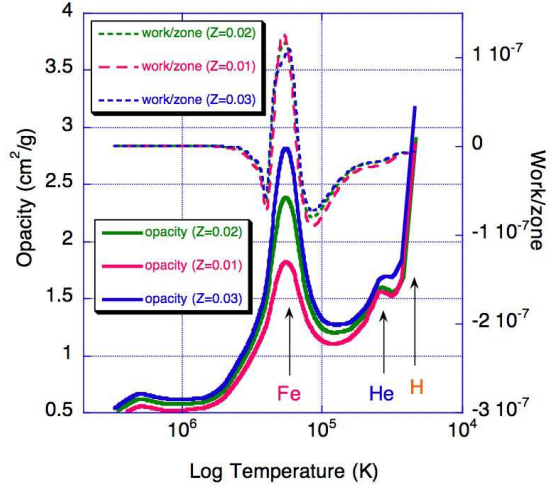}
\caption{For fixed effective temperature, radius ($\sim$23 R$_{\odot}$), luminosity, and Y abundance, the pulsation driving increases with increasing Z.}
\label{Fig12}
\end{figure}

\begin{figure}
\center
\includegraphics[width=\columnwidth]{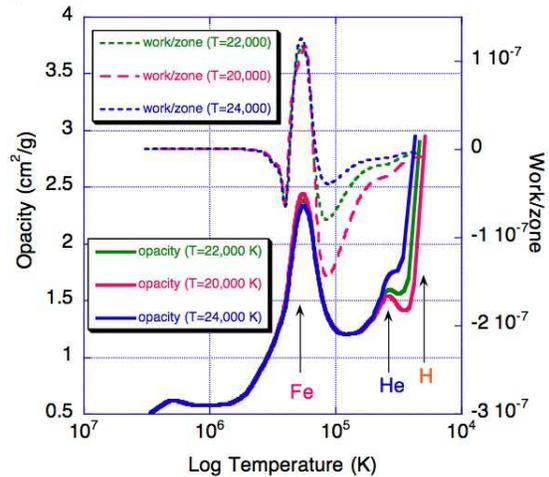}
\caption{For fixed composition and luminosity, linear growth rate increases for higher T$_{\rm eff}$ (younger) models.}
\label{Fig13}
\end{figure}

\begin{figure}
\center
\includegraphics[width=\columnwidth]{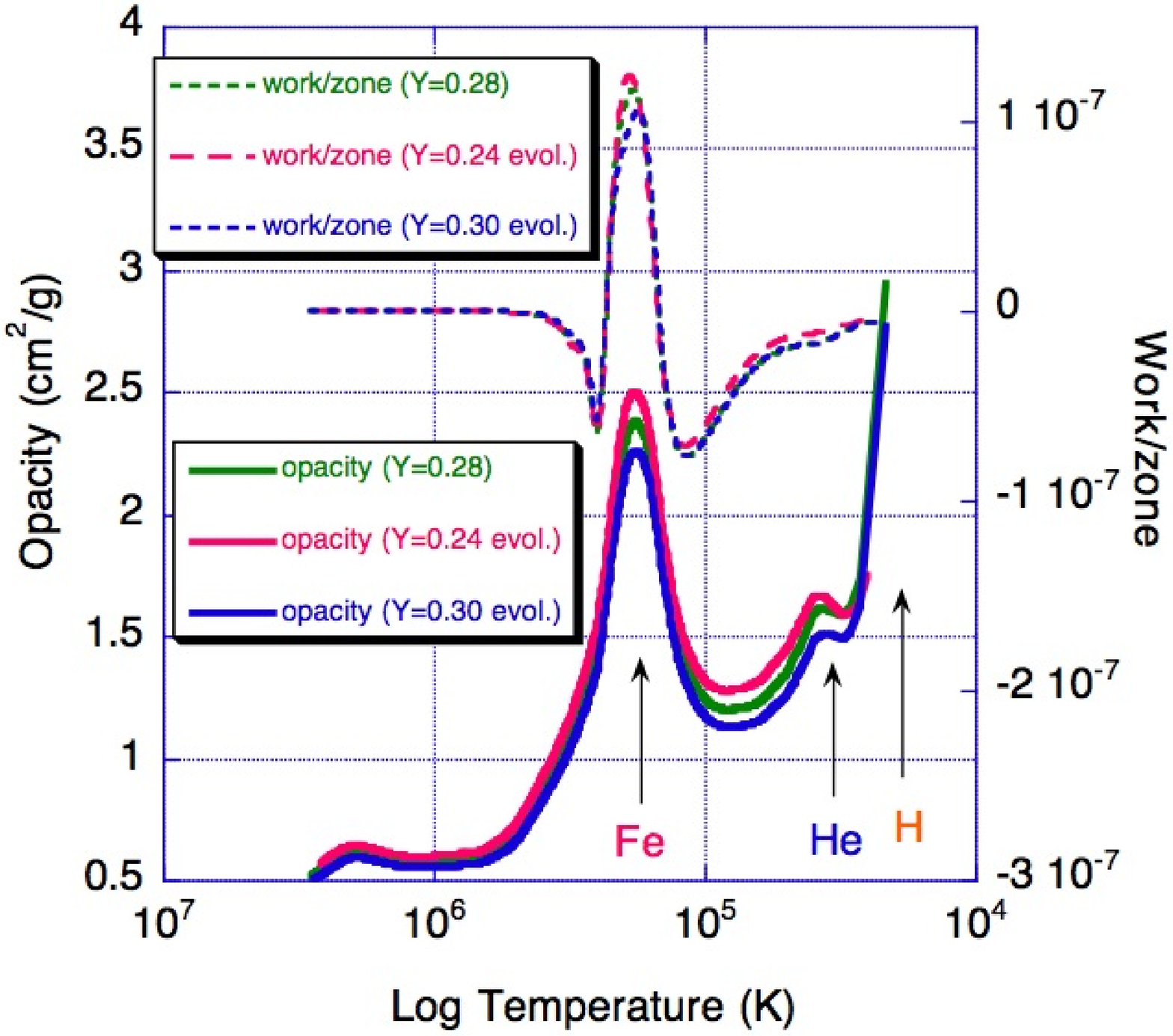}
\caption{Linear pulsation behavior of evolved 20 \msun\ models with varying initial Y and with T$_{\rm eff}$ = 22,139 K.}
\label{Fig14}
\end{figure}

In Figures \ref{Fig11}-\ref{Fig13}, we have only studied envelope models in which we artificially vary Y, T$_{\rm eff}$, and Z.  These models were made by evolving a model to a given point with one set of parameters, and then artificially holding all but one of the parameters constant while varying the last one.  For example, the different models in Figure \ref{Fig11} were made by holding radius, luminosity and temperature constant while varying the He abundance (Y).  However, when models are evolved from the ZAMS with the changed He abundance, the luminosity, radius, and envelope structure change relative to the original model.  In this case, the structure changes such that pulsation driving decreases with increasing He abundance (Figure \ref{Fig14}).

Tables \ref{Table3} and \ref{Table4} summarize the linear pulsation behavior of the models used in the parameter study in Figures \ref{Fig11}-\ref{Fig13}, and for the evolved models shown in Figure \ref{Fig14}.

\begin{table}
\caption{Parameter study for linear pulsations of models, varying Y, Z, and T$_{\rm eff}$.}
\begin{tabular}{llll}
\hline
	  &   &  	Fund. mode  &  	Growth rate/\\
Y, Z     &       T$_{\rm eff}$ (K)          &       period & period \\
           &   &  (days)  & (+$\equiv$unstable)\\
\hline
0.24, 0.02  &  22,000  &  0.888	  &  -0.0318\\
0.28, 0.02  &  22,000  &  0.877  &  	-0.0236\\
0.32, 0.02  &  22,000  &  0.868  &  	-0.0160\\
\hline
\\
0.28, 0.01  &  22,000	  &  0.833	  &  -0.0556\\
0.28, 0.02  &  22,000	  &  0.877	  &  -0.0236\\
0.28, 0.03  &  22,000  &  	0.916	  &  +0.0186\\
\hline
\\
0.28, 0.02	  &  24,000	  &  0.644	  &  +0.0374\\
0.28, 0.02	  &  22,000	  &  0.877	  &  -0.0236\\
0.28, 0.02	  &  20,000	  &  1.227	  &  -0.1028\\
\end{tabular}
\label{Table3}
\end{table}

\begin{table}
\caption{Linear pulsation behavior of 20 \msun\ evolved models with T$_{\rm eff}$ = 22,139 K.}
\begin{tabular}{llll}
\hline
	  &   &  	Fund. mode  &  	Growth rate/\\
Y, Z     &       T$_{\rm eff}$ (K)          &       period & period \\
           &   &  (days)  & (+$\equiv$unstable)\\
 \hline
 \\
			
0.24, 0.02  & 	98,988	  & 0.756  & 	-0.0166\\
0.28, 0.02	  & 111,370  & 	0.858  & 	-0.0189\\
0.30, 0.02	  & 125,015	  & 0.980	  & -0.0223\\

\end{tabular}
\label{Table4}
\end{table}

\begin{figure}
\center
\includegraphics[width=\columnwidth]{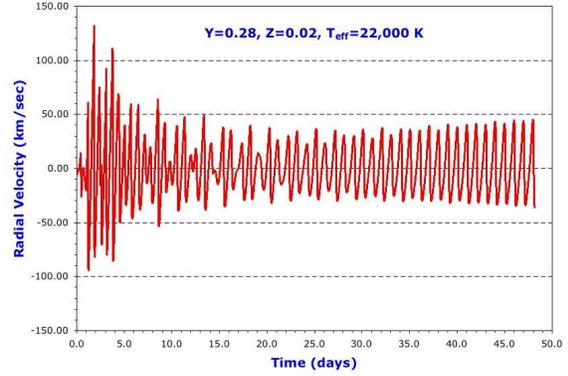}
\caption{Radial velocity of the photosphere vs. time for evolved 20 \msun\ model with Y=0.28, Z=0.02.  The amplitude rapidly increases and then decreases as the preferred period appears to be longer than the initial one, with slow amplitude growth thereafter.}
\label{Fig15}
\end{figure}

\begin{figure}
\center
\includegraphics[width=\columnwidth]{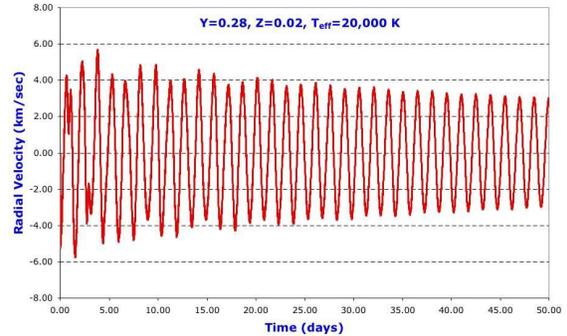}
\caption{Radial velocity of the photosphere vs. time for the same model as in Figure \ref{Fig15}, but with the T$_{\rm eff}$ changed to 20,000 K, while all other parameters are held constant.  The pulsation amplitude decreases with the decrease in effective temperature.}
\label{Fig16}
\end{figure}

\begin{figure}
\center
\includegraphics[width=\columnwidth]{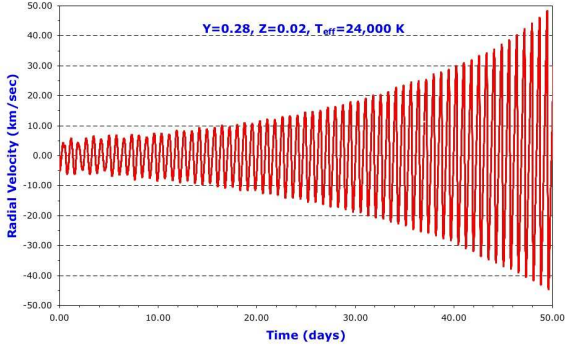}
\caption{Radial velocity of the photosphere vs.\,time for the model from Figure \ref{Fig15} with the temperature increased to 24,000 K.  The increase in effective temperature results in a growth in the pulsation amplitudes.}
\label{Fig17}
\end{figure}

\begin{figure}
\center
\includegraphics[width=\columnwidth]{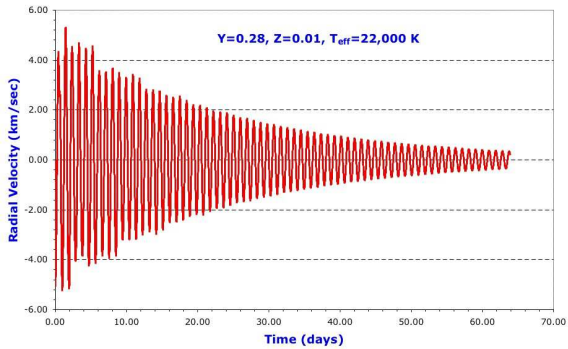}
\caption{Radial velocity of the photosphere vs. time for the model from Figure \ref{Fig15}, but with the metallicity changed to 0.01, and all other parameters held constant.  The pulsation amplitude damps quickly when the Z abundance is decreased.}
\label{Fig18}
\end{figure}

\begin{figure}
\center
\includegraphics[width=\columnwidth]{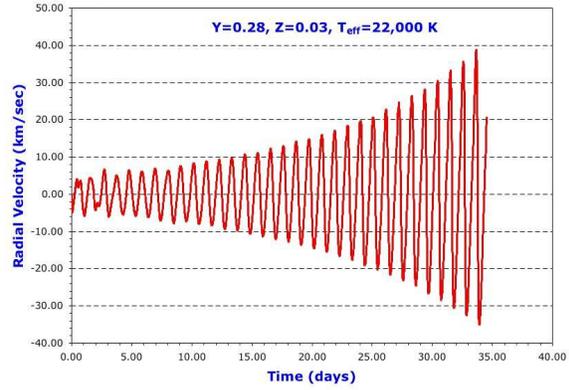}
\caption{As for Figure \ref{Fig18}, but with the metallicity increased to 0.03.  In this case, the pulsation amplitudes grow rapidly.}
\label{Fig19}
\end{figure}

\begin{figure}
\center
\includegraphics[width=\columnwidth]{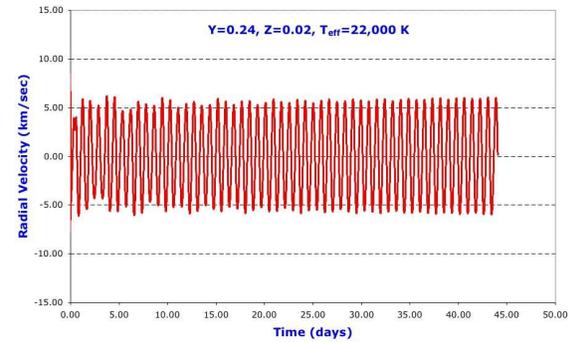}
\caption{Radial velocity of the photosphere vs. time for the model from Figure \ref{Fig15} with the helium abundance (Y) decreased to 0.24.  When Y is decreased in the envelope, the pulsation amplitude grows slowly, and the amplitude is smaller.}
\label{Fig20}
\end{figure}

\begin{figure}
\center
\includegraphics[width=\columnwidth]{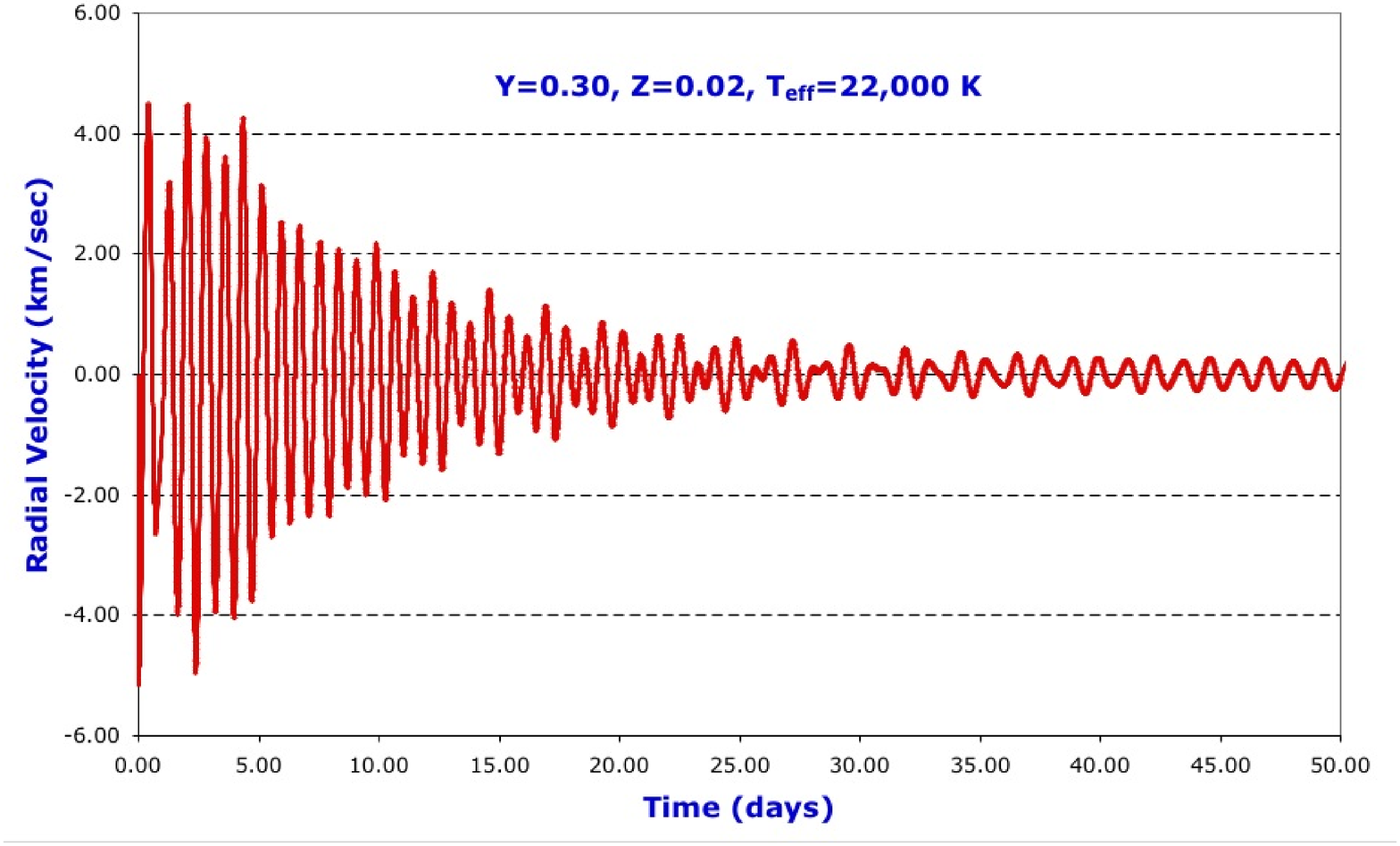}
\caption{As for Figure 20, but with the helium abundance increased instead of decreased.  The increase in helium abundance rapidly damps the pulsation amplitude.}
\label{Fig21}
\end{figure}

Next we show a series of plots (Figures \ref{Fig15}-\ref{Fig21}) of the behavior of these models in the nonlinear hydrodynamics code. Models are initiated in the radial fundamental mode with radial velocity amplitude 5 km/sec.

The abundance of Fe and He in stellar envelopes is predicted to affect the pulsational instability and amplitudes of 20 \msun\ models with photospheric temperatures near 22,000 K at the edge of the $\beta$  Cep instability strip.   The pulsations are driven by the $\kappa$ effect in the region of ionization of Fe-group elements near 200,000 K.  In our models, the pulsation driving and amplitude increase with increasing Z (increasing Fe) and decreasing Y (He) in envelope.  For the metallicity dependence, this behavior is consistent with the results of \citet{shiode12}.  

In future work, we could explore the effects of different masses, overtones, zoning, and depth of the envelope models.  We would also like to test predictions observationally.  However, stars of this mass are relatively rare, and evolve through this evolution state (shell H-burning) relatively rapidly, so may be difficult to observe.  In the longer lived main-sequence phase, these stars are nonradial pulsators, and rotating rapidly, whereas we have only considered radial and non-rotating models.  It would be useful to extend these models to 2D to include the effects of rotation.  Other improvements could include the ability to lose mass from the surface, allowing us to follow our models through an outburst and recovery.  

\begin{figure}
\center
\includegraphics[width=\columnwidth]{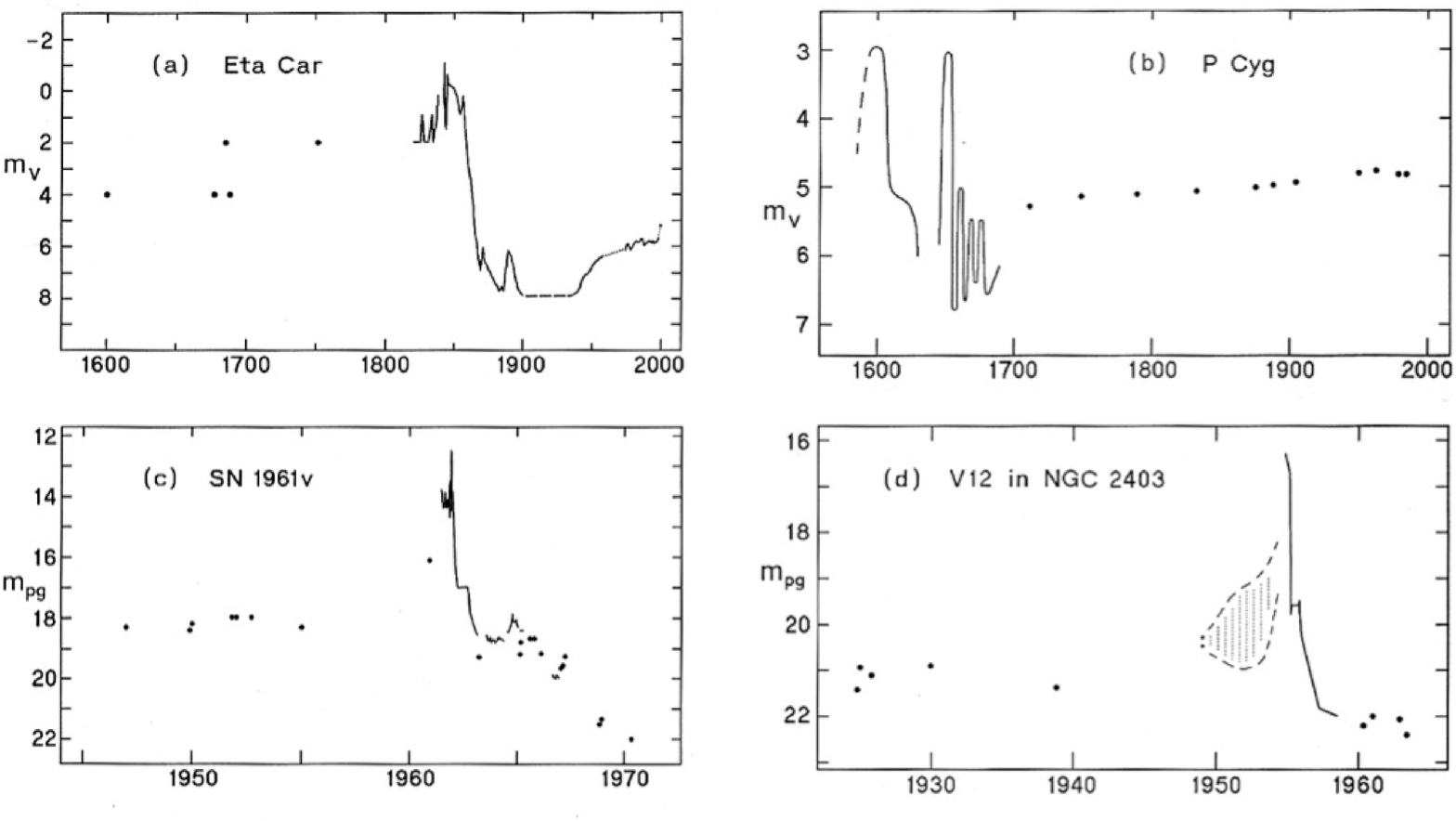}
\caption{Light curves of four supernova impostors.  a) Historical light curve of Eta Car.  b)  Historical light curve of P Cyg.  c) Light curve of SN1961v.  d)  Light curve of V12, an extragalactic LBV.  From \citet{humphreys99}.}
\label{Fig22}
\end{figure}

\begin{figure}
\center
\includegraphics[width=\columnwidth]{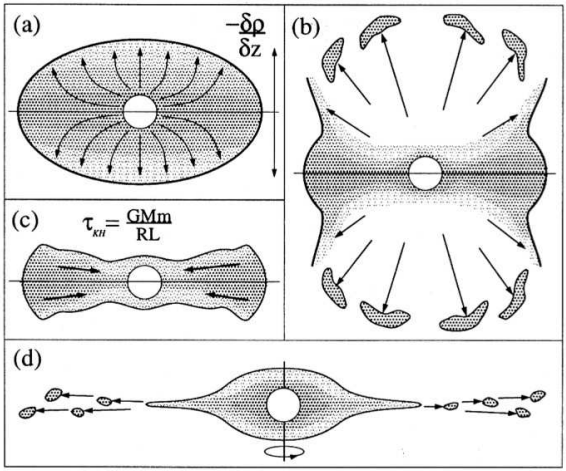}
\caption{Proposed scenario for Eta Car from \citet{smith98}.  A pre-eruption rotating star has a vertical density gradient.  In panel b, the Great Eruption ejects material from the poles of the star.  Post eruption, the stellar envelope readjusts itself (panel c).  During the 1890 eruption, material is ejected from the rapidly rotating envelope in the equatorial plane (panel d).}
\label{Fig23}
\end{figure}

\begin{figure*}
\center
\includegraphics[width=\columnwidth]{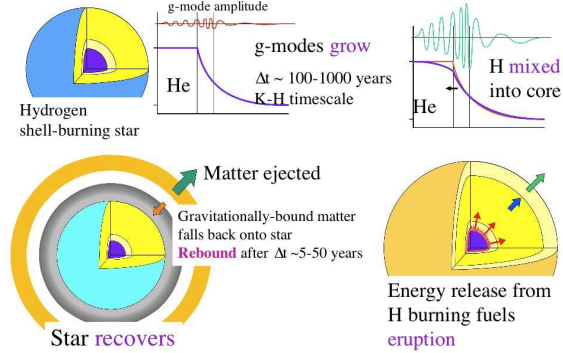}
\caption{Suggested scenario for initiating giant eruptions by recurring nuclear event.  The amplitude of gravity modes grows on a thermal timescale, mixing some hydrogen into the core.  This triggers a burst of nuclear energy, which causes material to be ejected.  This ejection is accompanied by a large increase in luminosity.  After the eruption, some of the material settles back onto the star and the cycle begins again.}
\label{Fig24}
\end{figure*}

\section{Giant Eruptions}

\subsection{Giant eruptions and rebounds}

As discussed by \citet{humphreys99}, there is evidence for several of the most luminous LBVs  (Eta Car, P Cyg, SN 1961v and V12 in NGC 2403) that have shown giant eruptions, that a giant eruption is followed by a `bounce' or `rebound' as evidenced by a smaller increase in luminosity, one or more decades later (Figure \ref{Fig22}).

Eta Carinae has been observed during two giant eruptions: the Great eruption in the 1840s, and a lesser eruption in the 1890s.  The Great Eruption lasted for around 20 years, and was accompanied by a brightening of about 2 bolometric magnitudes.  Measurements of the nebula produced by this event, the Homunculus, indicate that this eruption ejected on the order of 10 \msun.  The lesser eruption in the 1890s may have been a rebound from the Great Eruption, and ejected less mass, visible in the Little Homunculus; see Figure \ref{Fig23} outlining this scenario from \citet{smith98}.  There is also evidence for previous eruptions in Eta Car occurring about 200 years ago.  And there is some evidence that the eruption cycle in Eta Car is related to the orbital period of a binary companion \citep{smithfrew11}.

A second galactic LBV, P Cyg, was observed during a giant eruption, first around 1600 and again around 1640.  In this star, the eruptions are shorter, with a duration of about 6 years.  Again, the eruptions were accompanied by an increase in bolometric magnitude.  There is also evidence for previous eruptions between 900 and 2000 years ago.

The `rebound' time between the greater and lesser eruption is expected to be the thermal, or Kelvin-Helmholtz timescale of the matter involved in the outburst. 

The Kelvin-Helmoltz time is approximately
\begin{equation}
t_{K-H} \sim {\frac{GM^{2}}{RL}}, 
\end{equation}

\noindent
where $G$ is Newton's gravitational constant, $M$ is stellar mass, $R$ is stellar radius, and $L$ is stellar luminosity \citep[see, e.g.,][p. 26]{aerts10}.  We can use this formula to solve for the amount of mass above the core that must have participated in the eruption ($\Delta$$M$), and is recovering from the eruption on this timescale.

If we convert to solar units and years, and break the mass up into $M$ (total mass of the star) and $\Delta$$M$ (mass of the recovering layer), then we derive

\begin{equation}
\noindent
\Delta M = 3.2 \times 10^{-8}{\frac{RLt_{recov}}{M}}\,\, \rm{M}_{\odot}.
\end{equation}

If we take the time between the greater and lesser eruption for Eta Carinae as $t_{recov} \sim$30 years, and adopt for Eta Car $M$=80 \msun, $R$=200 R$_{\odot}$, $L$=5 $\times 10^6$ L$_{\odot}$, then we find that the amount of mass that must have participated in the eruption is $\Delta M$ = 12 \msun.

This large amount of participating mass, as well as the large luminosity change of Eta Car during outburst, imply a very deep-seated mechanism for the outbursts. The result also implies that, for a given amount of mass that escapes, several times more mass settles back to near its initial configuration during the recovery.

Because of the low mass in the outer layers of these luminous stars (recall that 99.8\% of the stellar mass lies within the inner several percent of the stellar radius), the origin of the giant eruptions ejecting tens of solar masses must originate deep in the core of star.  Considering the behavior of stars exhibiting giant eruptions such as Eta Car and P Cyg, a mechanism for the giant eruptions must turn on relatively suddenly, and then, perhaps after a bounce/rebound, turn off for hundreds or even thousands of years.  The mechanism must also generate additional energy, since large luminosity increases are seen during outbursts, and the mechanism needs to lift a large amount of mass out of a deep gravitational potential well.

\subsection{Gravity-mode mixing} 
 
Here we outline a scenario that might produce the conditions described above, namely a mixing episode, perhaps initiated by gravity-mode oscillations, that mixes additional hydrogen inward to a region where it burns abruptly.

Gravity modes can be initiated by turbulent processes in the envelope \citep[e.g.,][]{stothers00,samadi10}.  For evolved stars, a dense spectrum of g modes is present with closely spaced nodes in a g-mode cavity formed at the edge of a burned-out core where there is a steep composition gradient \citep[see, e.g.,][for examples based on shell hydrogen-burning $\delta$ Scuti stars]{guzik00,templeton00a,templeton00b}. While g modes oscillate at periods near the dynamical timescale (days), they grow (or damp) in amplitude over the thermal or Kelvin-Helmholtz timescale ($\sim$100 years for Eta Car; $\sim$1000 years for P Cyg).  The g-mode amplitudes could grow in the composition gradient region around the H-burning shell, and cause some hydrogen to mix into the H-exhausted core, producing a burst of nuclear energy and triggering an eruption, accompanied by a large luminosity increase and ejection of some mass.  Gravitationally-bound material then settles back on to the star.  The recovery time should be the Kelvin-Helmholtz timescale for the infalling material, 5-50 years, or the recovery times seen between major outbursts in Eta Car.  The g modes are initiated again later to slowly grow in amplitude, restarting the cycle.  This process is illustrated in Figure \ref{Fig24}.

\subsection{Additional possible mechanisms for initiating giant eruptions}

Here we mention a few other mechanisms that may operate deep enough in the star and involve enough of the stellar mass to initiate giant eruptions, and should also be explored.

\subsubsection{Secular/thermal instability}

In standard pulsation analysis, such has been described above, the pulsation frequencies are found by perturbing a model in hydrostatic equilibrium.  The resulting periods are on the order of the dynamical timescale of the star.  However, an analysis could be done where the model is forced to be in hydrostatic equilibrium, but oscillations around thermal equilibrium, on the longer Kelvin-Helmholtz timescale are followed.  The theory of secular stability and application to stellar structure and pulsations is reviewed by \citet{hansen78}.  Pulsations arising from such secular or thermal instabilities (if found), may have the right periods and depths in the star to initiate the giant eruptions. 

\subsubsection{Epsilon mechanism}

The epsilon mechanism is a pulsational instability arising from modulated nuclear energy generation in the cores of stars \citep{kippenhahn90}.  The mechanism continues to be explored to explain pulsations in massive stars \citep[e.g.,][]{moravveji11,shiode12} and in low-mass low metallicity stars \citep[e.g.,][]{sonoi12}.  However, modes initiated by this mechanism may also be damped by turbulent viscosity or processes in the overlying stellar layers, and so far no variable stars classes have been been verified observationally with pulsations arising from this mechanism.  Since this instability would originate deep in the core of an LBV star where nuclear energy generation is occurring, it is in the right location to play a role in initiating a giant eruption.

\subsubsection{Standing Accretion Shock Instability}

The SASI (Standing Accretion Shock Instability) mechanism \citep[see, e.g.,][]{foglizzo08,blondin03,scheck08} has been suggested to assist in explosions in core-collapse supernova, and perhaps also explains the asymmetry of the ejecta.  Accretion shocks occur not only in supernovae, but also in star formation and accreting compact objects.  This instability does not necessarily require conditions present only in core-collapse supernovae, such as neutrino heating, but requires inflow and vorticity plus acoustic waves, and produces a nonradial $l$=1 oscillation that can grow in amplitude.  The mechanism was even demonstrated in a scaled laboratory experiment using shallow water \citep{foglizzo12}.  Since for LBV stars a longer timescale is available for the instability to grow compared to the time after the post-bounce accretion phase of a supernova, perhaps this mechanism could also play a role in assisting the LBV giant eruptions.

\subsubsection{Tidal effects from binary companions and rapid core rotation}

The roles of tidal forces from binary companions and possible rapid core rotation in initiating LBV giant eruptions should also be investigated.

\section{Future work and requirements}

In this paper, we have only investigated in 1-dimension (1-D), for radial modes, in spherically symmetric, single, non-rotating models, some effects of helium, metallicity, effective temperature, and the luminosity/mass ratio on linear and nonlinear pulsations, and we have shown that taking into account the time-dependence of convection tin he hydrodynamic models allows the Eddington limit to be exceeded in the deeper envelope layers, and may initiate mass outflows.

There is utility in extending and improving these 1-D linear and nonlinear hydrodynamical models.  We could refine zoning, try to improve the numerical stability by eliminating opacity table interpolation, continue running the models for longer times to look for possible mode changes, extend the parameter studies to additional masses, metallicities, and study the effect of the time-dependent convection parametrization.  

However, other aspects of these problems seem more important for understanding LBVs, and require advances in modeling.  It will be important to add an outflow boundary condition to the evolution and pulsation models to model the most massive stars that are not in hydrostatic equilibrium in their outermost layers.   It may be important to numerical stability and keeping reasonable timesteps to use an adaptive grid code to rezone the envelope to adapt to the changes in radius during the pulsations.  A means to remove and replenish mass in the envelope and follow mass loss is critical to predicting mass-loss rates or possible recovery from a mass-ejection episode.  For this improvement, one may need to deal with opacity/EOS for low-density, non LTE effects, and improve radiation transport models.

In addition, not only radial, but also nonradial modes likely play a role, and so a nonlinear and nonradial pulsation code may be needed.  Nonradial phenomena are evident in many LBV stars \citep{weis12} such as the bipolar nebula and `skirt' in Eta Car.

Two-dimensional and three-dimensional codes may be required to model the effects of convection and convective overshoot, and model the time dependence of convection more realistically, deal with rapid and probably differential rotation, the tidal effects of possible binary companions, or investigate episodic mixing that might be produced by gravity modes or convective plumes.

The physics models and computational capabilities to handle these diverse phenomena are available in astrophysical codes being developed today, but perhaps not assembled to address specific aspects of LBV pulsation and mass outflow or interior instability problems in any one code \citep[see, e.g.,][for a summary of results from some recent stellar interior hydrodynamics codes.]{guzik10}.  Some relevant capabilities were available in almost forgotten codes, such as the \citet{bowen88} code developed to handle Mira mass loss, or the adaptive grid hydro code of \citet{gehmeyr93} applied to RR Lyrae variable star pulsations.  We hope that some of these tools can be adapted or resurrected and applied to the yet unexplained phenomena in LBV stars.

\section*{Acknowlegements}
The authors acknowledge Arthur N. Cox, Dale Ostlie, Kate Despain, Jay Onifer, Michael Soukup, Matthew Templeton, Benjamin Austin, Dean Pesnell, Brandon Peterson, and Siobahn Morgan, for calculating countless LBV models, attending and presenting at numerous conferences, and developing and providing the computational tools over the past 20 years.  The authors thank many colleagues including Paul Bradley, Roberta Humphreys, Kris Davidson, S. Shore, H. Lamers, N. Smith, B. Wolf, A. Heger, A. Maeder, S. Owocki, N. Langer, and W. Glatzel for encouragement and important discussions. The authors also thank the anonymous referee who provided a thorough review of this paper with a short deadline despite its preliminary draft form when submitted.  This work was performed for the U. S. Department of Energy by Los  Alamos National Laboratory under Contract No. DE-AC52-06NA2-5396.

\end{document}